\documentclass{aa}

\usepackage[varg]{txfonts}
\usepackage{graphicx}
\usepackage{array, booktabs, makecell}
\usepackage{siunitx}
\usepackage[version=4]{mhchem}
\usepackage{textcomp, gensymb}
\usepackage{amsmath}
\usepackage{amsfonts}
\usepackage{comment}
\usepackage{multicol}
\usepackage[utf8]{inputenc}
\usepackage[english]{babel}
\usepackage{natbib}
\bibpunct{(}{)}{;}{a}{}{,}
\usepackage[hidelinks,colorlinks=true,linkcolor=blue,citecolor=blue]{hyperref}
\usepackage[dvipsnames]{xcolor}

\newcommand{\be}{\begin{equation}}
\newcommand{\ee}{\end{equation}}
\newcommand{\bea}{\begin{eqnarray}}
\newcommand{\eea}{\end{eqnarray}}

\begin{document}

\title{AGN hosting jets. I: A semi-analytical model for the evolution of radio galaxies}
\author{Pau Beltrán-Palau \inst{1}, Manel Perucho \inst{1,2}, Jos\'e Mar\'{\i}a Mart\'{\i}\inst{1,2}}

\institute{\inst{1} Departament d’Astronomia i Astrofísica, Universitat de València, C/ Dr. Moliner, 50, 46100, Burjassot, Val\`encia, Spain \\
\inst{2} Observatori Astronòmic, Universitat de València, C/ Catedràtic José Beltrán 2, 46980, Paterna, Val\`encia, Spain
}

\date{Received xx / Accepted yy}

\abstract
{Three-dimensional simulations of relativistic jets are a useful tool to understand the evolution of jets and radio galaxies in detail.
However, computationally demanding as they are, their use is limited to a relatively small number of representative cases. When comparing to the distribution of large samples of objects in the luminosity-distance plane ($P-D$ plane), the most efficient approach is to use analytical or semi-analytical models that reproduce the evolution of the main parameters governing the dynamics and radio luminosity of the sources.}
{Our aim is to build a semi-analytical model that allows us to produce mock samples of radio galaxies to be compared with real populations and use this approach to constrain the general properties of active galaxies with jets in a cosmological context.}
{In this work, we present a new model for the evolution of radio galaxies based on the resolution of ordinary differential equations and inspired both by previous experience on numerical simulations of jets across several orders of magnitude in power, and by observational evidence.}
{Our results show remarkable agreement between the results given by the semi-analytical model and those obtained by both 2D and 3D relativistic hydrodynamics simulations of jets ranging from $10^{35}$~W to $10^{39}$~W. From the derived trajectories of powerful radio galaxies through the $P-D$ diagram (powers greater than $10^{36}$~ W), our model agrees with typical lifetimes of galactic activity of $\leq 500$~Myr. We also compare our results with previous models in the literature. In a follow-up paper, we use this model to generate mock populations of radio galaxies at low redshifts and compare them to the LoTSS sample.}
   {}

\titlerunning{AGN hosting jets. I: A semi-analytical model for the evolution of radio galaxies}
\authorrunning{Beltr\'an-Palau, Perucho \& Mart\'{\i}}
\maketitle

\section{Introduction}

Galactic activity is known to be triggered by accretion on supermassive black holes at the host's nucleus \citep[AGN, see, e.g.][]{2013peag.book.....N}. This process can result in the formation of relativistic jets with different powers \citep{1984RvMP...56..255B}. Their relevance to galactic and cluster evolution is obvious, since AGN in general, and relativistic jets and outflows in particular, have been shown to play a relevant role 
in terms of feedback and heating of the interstellar and intergalactic media \citep[ISM and IGM, respectively;][]{2007ARA&A..45..117M,2012ARA&A..50..455F}. 

Powerful relativistic jets such as those found in FRII radio galaxies \citep{1974MNRAS.167P..31F,1984RvMP...56..255B, 2019ARA&A..57..467B} generate collimated, supersonic outflows and produce bright hotspots at the impact region with the ambient medium, feeding extended radio lobes made of the back-flowing plasma processed at the jet terminal shocks (the hotspots). The lobes are the observational counterpart of the "cocoons" appearing in numerical simulations, so we will use these two terms, lobes and cocoons, as synonyms throughout the text.

In the case of lower power jets, they can be decelerated by mass entrainment, becoming subsonic and developing plumed, irregular structures \citep[FRI radio galaxies, see, e.g.][]{2002MNRAS.336.1161L,2014MNRAS.437.3405L}. The physical processes and spatial/temporal scales in which these jets perturb the ambient medium may range from mixing of buoyant, hot bubbles with the ISM within few kiloparsecs of the AGN for the weakest jets to shock heating up to hundreds or even thousands of kiloparsecs for the powerful ones \citep[e.g.,][]{2003MNRAS.344L..43F,2024Natur.633..537O}. 

Radio galaxies are, among AGN, those that show misaligned radio jets with respect to the line of sight, and kiloparsec scale lobes that can be observed at low frequencies. These structures are known, after theoretical and numerical modelling, to evolve through the ISM and IGM with different temporal scales and properties that depend on the initial jet power and ambient density/pressure profiles \citep[see, e.g.,][]{2007MNRAS.382..526P,2019MNRAS.482.3718P,2022MNRAS.510.2084P,2013MNRAS.430..174H,2016MNRAS.461.2025E,2016A&A...596A..12M,2019A&A...621A.132M,2022A&A...659A.139M,2021MNRAS.508.5239Y,2021ApJ...920..144S}.

On the one hand, numerical simulations of relativistic outflows have proven to be an excellent tool to study the large scale evolution and properties of radio galaxies. However, they require significant computing time, which invalidates their usage to explore a vast parameter range. On the other hand, analytical or semi-analytical models have succeeded in reproducing the evolution of simulated jets \citep[e.g.,][]{2011ApJ...743...42P,2019MNRAS.482.3718P,2018MNRAS.475.2768H}. Therefore, albeit with limitations, such models can be used to explore wider regions of the parameter space, which is interesting to understand  the evolution of large samples of radio galaxies. This approach can eventually allow us to compare with observational databases and to study the cosmological evolution of galactic activity. 

Previous analytical works on jet evolution have been mainly guided by the self-similar evolution model presented in \citet{1997MNRAS.286..215K}, superseding previous models for uniform  \citep{1974MNRAS.166..513S} or density-decreasing ambient media \citep{1982IAUS...97...21B,1987MNRAS.226..531G}. \citet{1997MNRAS.286..215K} assumed initially ballistic jets evolving 1) through a decreasing density ambient medium, 2) with a constant opening angle, 3) reaching pressure equilibrium with their direct environment (the cocoon region) after recollimating, and 4) triggering a strong shock (the hotspot) at the impact region with the ambient medium. The model assumes an adiabatic expansion of the shocked gas from the hotspot to the lobes and a self-similar expansion of the shocked region (leading to an approximately constant pressure factor between the hotspot and the lobe).

Another set of models were devoted to study the evolution of radio luminosity with distance/time in radio sources \citep[][]{1997MNRAS.292..723K,1999AJ....117..677B,2002A&A...391..127M}. These models assume that all jet particles emit radiation and are in equipartition with the magnetic field, and take into account the energy loss due to work exerted in the lobe expansion. In \citet{2006PhDT........38B,2006MNRAS.372..381B,2007ApJ...658..217B,2008ApJ...682L..17B}, the authors test the aforementioned dynamical and radiative models against the properties of complete radio surveys such as 3CRR, 6CE, and 7CRS. \citet{2006MNRAS.372..381B,2007ApJ...658..217B} used a Monte Carlo simulation on jet power, keeping the host galaxy properties fixed in all the simulated radio sources. \citet{2000A&A...362..447K} presented an improved version of the \citet{1997MNRAS.292..723K}
model, in which a three-dimensional distribution of the emissivity was calculated, taking into account the backflow velocity and the energy losses of the relativistic particles. However, this model is intended to be applied to individual sources and requires input from radio observations of jet lobes.

In the case of powerful radio galaxies, numerical simulations have allowed to test the validity of analytical, simplified models and, to a large extent, have shown that during the strong shock expansion phase, a model such as Begelman-Cioffi's \citep[][BC, from now on; see also \citet{1990ApJ...355..416D} for a similar model]{1989ApJ...345L..21B} extended to allow for variable ambient density and jet head velocity \citep[eBC model, from now on;][]{2002MNRAS.331..615S,2007MNRAS.382..526P} can reproduce the evolution of the global lobe parameters \citep{2011ApJ...743...42P,2019MNRAS.482.3718P,2022MNRAS.510.2084P}. 

In \citet{2018MNRAS.475.2768H}, the author presents a dynamical model based on results from FRII jet evolution simulations in which the basic assumptions are: constant power injection, light and relativistic outflows, shock-driven expansion, isothermal ISM/IGM atmosphere, constant fraction of jet energy flux converted into lobe pressure and radiating particles, pressure balance within the shocked region, Rankine-Hugoniot conditions valid at the shocks, lobe axial size equal to the radial position of the shock, and adiabatic expansion. The dynamical model is conceptually similar to the eBC one, although it takes into account the jet thrust to compute the head advance in order to fix the conditions to solve the evolution differential equations for lobe length and width independently. This is in contrast to the eBC model, which requires fits to the time evolution function for the advance velocity to compute the rest of values \cite[e.g.,][]{2011ApJ...743...42P}. The comparisons with numerical simulations show the robust predictive power of such simple models. 

\citet{2018MNRAS.475.2768H} also includes the calculation of the radiative output in his work, assuming a purely jet non-thermal composition and sub-equipartition between the magnetic field and the particles. The model includes corrections to account for adiabatic and radiative losses and the resulting changes in the particle distribution. According to this work, the distribution of source lifetimes is a critical parameter determining the size of radio galaxies, but not so much for luminosity. Another relevant conclusion of this work is the rapid decay in luminosity once jet ejection ends. 

\citet{2015ApJ...806...59T} and \citet{2023MNRAS.518..945T} presented more elaborated models that included a two-dimensional approach to lobe expansion, assuming cylindrical symmetry. Furthermore, the authors extend the evolutionary models to low-power radio galaxies and to the regime of lobe-ambient pressure equilibrium. In \citet{2023MNRAS.518..945T} the authors added the jet contribution via a spine-sheath modelisation and the calculation of its contribution to the simulated source synchrotron luminosity. The luminosity is computed by a combination of a Lagrangian particle approach and the particle Lorentz factor estimates from \citet{1997MNRAS.292..723K}. This model, named RAiSE, was shown to be able to reproduce the evolution of simulated radio galaxies by means of relativistic hydrodynamics simulations of both low-power and powerful jets \citep{2022MNRAS.511.5225Y}. 

In this paper, we present a model that has a slightly larger degree of complexity than that in \citet{2018MNRAS.475.2768H}, but smaller than those presented in \citet{2015ApJ...806...59T,2023MNRAS.518..945T}. 
The model is simulation-motivated and based on the BC model and its extensions \citep[][]{1989ApJ...345L..21B,2002MNRAS.331..615S,2007MNRAS.382..526P}, to which we add the radiative model based on the one developed by \citet{1997MNRAS.292..723K}.

In comparison with \citet{2015ApJ...806...59T,2023MNRAS.518..945T}, our model is simpler and uses a different prescription to compute the radiative output. Our model includes the possibility of a thermal population, jet mass-load, and we use a different approach to the evolution of jet luminosity beyond the lobe overpressure phase. Our dynamical model also separates the axial and sideways expansion of the lobe, thus removing the self-similarity imposed in \citet{1997MNRAS.286..215K}. 

On the one hand, the consideration of a thermal population allows us to alleviate the role of emitting particles in driving the expansion and, therefore, the losses this would imply. On the other hand, the independent calculation of radial and axial evolution allows us to account for the extra adiabatic losses suffered by the emitting particles with respect to self-similar models, as pointed out by \citet{1997MNRAS.292..723K,1999AJ....117..677B,2002A&A...391..127M}. Furthermore, we add a new method to evaluate the radio-luminosity of decelerated and decollimated low-power sources, in which particle acceleration cannot take place at the hotspot, but along the jet. This is another novel approach in our model, allowing us to follow the luminosity evolution of sources across a wide range of jet powers, beyond the strong shock regime and into the phase in which the lobe is in pressure equilibrium with the ambient medium.

Our aim is to apply this model informed by numerical simulations to the study of active galaxy populations, following the idea by \citet{2006MNRAS.372..381B,2007ApJ...658..217B}, and to extend it to a wider range of jet powers to compare with large samples of radio galaxies, such as LoTSS \citep{2019A&A...622A..12H}. 

The paper is organized as follows. In sections~\ref{Dynamics} and \ref{Rlum}, we present our model for the dynamical and radio emission evolution of the sources, respectively. In sec.~\ref{Param}, we specify the parameter space that we use for the calculations of radio galaxy evolution. We present our results for different jet powers and environments in sec.~\ref{results}. These results are put into context and discussed in sec.~\ref{Discussion}. {Finally, sec.~\ref{conclusions} contains a brief summary of the present work.}

\section{Evolution model: Dynamics}
\label{Dynamics}

\subsection{Host galaxies}\label{sec:host1}

We model the ambient density profile with a spherically symmetric double King profile \citep[e.g.,][]{2002MNRAS.334..182H}, also used in the numerical simulations in which the model is based \citep[e.g.,][]{2011ApJ...743...42P,2019MNRAS.482.3718P}, taking into account both the galaxy and the group/cluster profiles, and a constant density for the gas between galaxies or clusters. Then, the density profile is given by
\be\label{density}
\rho(l)=\rho_{g,0}\left(1+\left(\frac{l}{a_g}\right)^2\right)^{\beta_g/2}+\rho_{c,0}\left(1+\left(\frac{l}{a_c}\right)^2\right)^{\beta_c/2}+\rho_{\rm IGM},
\ee
where $\rho_{g,0}$ and $\rho_{c,0}$ are the central densities of the galaxy and the cluster respectively, $\rho_{\rm IGM}$ is the density of the intergalactic or intercluster medium –depending on the mass resulting from the profile–, $a_g$ and $a_c$ are the corresponding radii of the galactic and cluster cores, respectively, $l$ is the distance from the galactic nucleus and $\beta_g$ and $\beta_c$ are the negative exponents that regulate the decreasing rate. These exponents cannot be lower than $-2$ in order to ensure that the jet shock is formed \citep{10.1093/mnras/250.3.581}. Note that, by considering spherical symmetry, we implicitly assume that our galaxies are located at the center of the clusters.

\begin{figure}[htbp]
\begin{center}
\includegraphics[width=90mm]{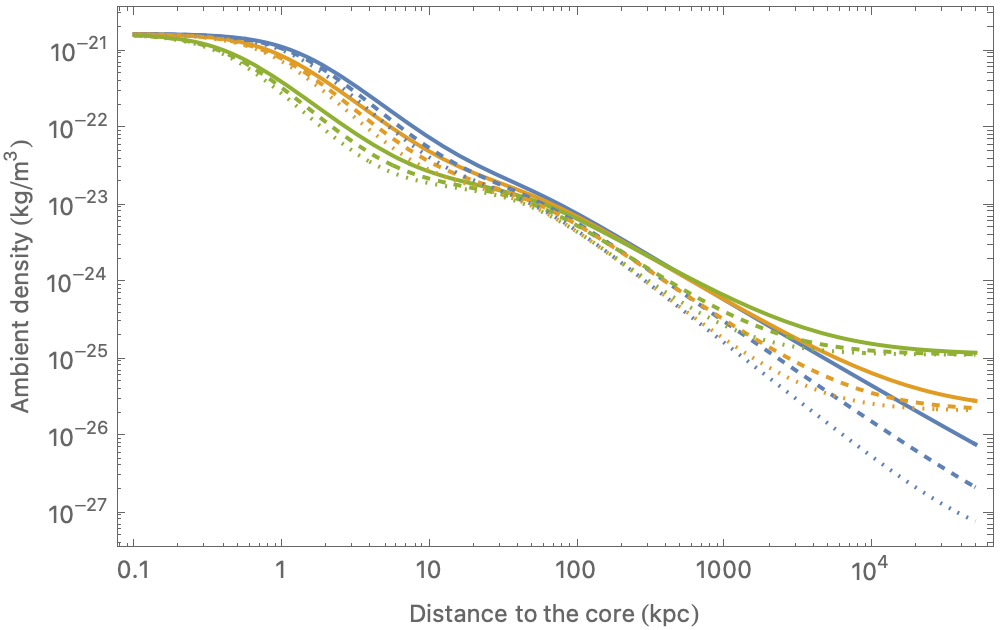}
\end{center}
\caption{Ambient density profiles for different redshifts ($z=0$ in blue, $z=3$ in orange and $z=6$ in green) and for different values of the exponents $\beta$ ($\beta_g=-1.6$ and $\beta_c=-1.1$ in solid curves, $\beta_g=-1.8$ and $\beta_c=-1.3$ in dashed curves and $\beta_g=-2$ and $\beta_c=-1.5$ in dotted curves). The radius of the core at $z=0$ are fixed at $a_{g,0} = 1.25\text{ kpc}$ and $a_c = 45\text{ kpc}$, and the central densities at $\rho_{g,0} = 1.673 \times 10^{-21} \text{ kg/m}^3$ ($\approx 1\text{ proton/cm}^3$) and $\rho_{c,0} = 1.673 \times 10^{-23} \text{ kg/m}^3$ ($\approx 0.01\text{ proton/cm}^3$).}
\label{rho_figures}
\end{figure}

A redshift dependence of the galactic core size and the IGM density are added to account for the cosmological evolution of the host galaxies. The galactic core is thus modeled according to the following expression \citep{2017MNRAS.469.4083S}
\be \label{eq:core}
a_g= \begin{cases}a_{g,0}, & \text { if } z \leq 2 \\ a_{g,0}[(1+z) / 3]^{-1.25}, & \text { if } z>2\end{cases},
\ee
where $a_{g,0}$ is the radius at $z=0$. On the other hand, the IGM density can also be taken as redshift dependent in order to account for the cosmological expansion \citep[e.g.,][]{2018A&A...617L...3B}: $\rho_{\rm IGM} =(3.345 \times 10^{-28} \mathrm{~kg} / \mathrm{m}^3)(1+z)^3$. In Fig. \ref{rho_figures} we represent the density profile for different values of $\beta_g$, $\beta_c$ and the redshift. The ambient medium is considered to be isothermal, which fixes the ambient pressure profile to be the same as for the density.

%
%

\subsection{Shock advance}\label{shock_advance}

The jets drive shocks into the host's ambient medium while the expansion velocities of the perturbations they trigger are supersonic. In this section we describe the methodology applied to solve the equations that describe this phase. Once the expansion becomes transonic, it proceeds at the sound speed of the ambient medium, and this phase is treated separately in the next section. 

As in the original BC model, the shocked region is modeled as cylindrical, with radius $r$ and length $d$. By imposing the one-dimensional Rankine-Hugoniot jump conditions between the cocoon and the ambient medium, we obtain a set of differential equations for the sideways propagation of the shock, which can be reduced to a single one 
\be\label{edo_r}
\dot r(t)^2=\frac{1}{2 {\rho_a(t)}}\left[(\Gamma+1) p_c(t)+(\Gamma-1) p_a(t)\right],
\ee
where the dot represents the derivative with respect to time, $r$ is the radial position of the shock, and $\Gamma$ is the adiabatic index of the gas, which we consider to be equal to $5/3$ both in the cocoon and in the ambient medium, assuming that the (non-relativistic) thermal population dominates the dynamics of the source. Finally, $p_c$ and $p_a$ are, respectively, the cocoon (post-shock) and ambient (pre-shock) pressures, and $\rho_a$ is the ambient density. Large post-shock sound-speeds ensure the homogenization of $p_c$, the pressure inside the shocked region, as shown by numerical simulations \citep[e.g.,][]{2014MNRAS.441.1488P,2019MNRAS.482.3718P}. The ambient pressure can be related to the ambient density by $p_a=\rho_a c_{s, \, a}^2/\Gamma$, with $c_{s, \, a}$ the (constant) sound speed in the isothermal ambient medium. 

For the radial expansion, we choose a particular point at the shock surface in order to fix the pre-shock density and pressure. The selected point is the middle point of the length of the cylinder, located at a distance $\sqrt{r^2+(d/2)^2}$ from the center of the galaxy. Therefore, in \eqref{edo_r} we take $\rho_a(t) = \rho(\!\sqrt{r(t)^2+(d(t)/2)^2})$. By doing this, we approximate the radial expansion of the cylinder against a changing ambient medium along its surface, which is represented by values at the aforementioned point.

To make the differential equation \eqref{edo_r} solvable, we need to inform it with two functions of time: the cocoon pressure $p_c(t)$ and the length of the cylinder representing the shocked region, $d(t)$, which enters Eq.~\eqref{edo_r} through the definition of $\rho_a(t)$. In our model, the evolution of the cocoon pressure follows from long-term numerical simulations of jets, as described below. Additionally, it is used to estimate the length of the shocked region taking into account that, in the case of relativistic jets, a significant amount of the total energy injected by the jet into the cocoon up to time $t$ ($L_0 t$, $L_0$ being the kinetic power of the jet, assumed constant) is converted into internal energy and hence
$p_c V = f L_0 t$, where $V$ is the cocoon volume at time $t$ and $f$ is the fraction of the original energy injected by the jet converted into internal energy in the cocoon. According to \cite{2017MNRAS.471L.120P}, $f\simeq 0.4$ for relativistic jets \citep[see also][]{2016MNRAS.461.2025E}. Therefore, the length of the shocked region can be estimated as
\be\label{marina}
d(t)\simeq \frac{f L_0 t}{p_c(t) \pi r(t)^2}.
\ee
With this, differential equation \eqref{edo_r} becomes
\be\label{edo_r2}
\dot{r}(t)^2=\frac{(\Gamma+1) p_c(t)}{2 \rho \left(\sqrt{r(t)^2+(\frac{f L_0 t}{2 p_c(t) \pi r(t)^2})^2}\right)}+\frac{\Gamma-1}{2 \Gamma}c_{s, \, a}^2.
\ee

We apply this equation in 5 distinct phases, defined by the critical times corresponding to the instants when $\sqrt{r^2+(d/2)^2}$ reaches the following values: $l_{\rm crit} = (a_g, l_{gc}, a_c, l_{IGM})$. We define $l_{gc}$ as the distance at which the cluster density (second term in \eqref{eq:core}) starts to dominate over the galactic one (first term), that is $\simeq a_g(\rho_c/\rho_g)^{1/\beta_g}$. Analogously, we define $l_{\rm IGM}$ as the distance at which the density reaches $\rho_{\rm IGM}$, i.e., $\simeq a_c(\rho_{\rm IGM}/\rho_c)^{1/\beta_c}$. Then, the first phase corresponds to the core region, the second to the region beyond the core where the galaxy's density still dominates over that of the cluster, the third to the region where the cluster dominates, the fourth to the region beyond the cluster's core radius, and the fifth to the constant density intergalactic/intercluster medium.
For each case, a different exponent $b$ for pressure evolution, $p_c \sim t^b$, is imposed, as informed by numerical simulations \citep{2011ApJ...743...42P,2019MNRAS.482.3718P}, which show that these exponents are very similar for all simulated jets, and as justified below. We will refer as $b_i$, with $i=1,...5$, to the exponents for each phase.

Once the values for the pressure evolution exponents are given, we need to provide initial conditions to solve Eq.~\eqref{edo_r2}. We start the radio galaxy evolution after 1000 years from the jet initial injection from the active nucleus. Using the same approach as in \citet{2024A&A...684A..45P} to estimate the values of post-shock pressure and shock length at the initial condition $t_0$, we assume that the source expands self-similarly before $t_0$, which has been reported to be a good approximation for hotspots/lobes within the inner kiloparsec \citep{2002ApJ...568..639P,2008ApJ...687..141K,2017Ap&SS.362...87O}. The cocoon's aspect ratio, $R_c = d/r$, is thus constant during this period (although it is taken to depend on the jet power). 

Considering that in this initial phase the cocoon pressure, $p_c$, evolves as $t^{b_1}$ and imposing the relation \eqref{marina}, we find that both $r$ and $d$ evolve as $t^{(1-b_1)/3}$. Therefore, in this initial phase $r(t)=r_0(t/t_0)^{(1-b_1)/3}$, where $r_0$ is the initial radius that will be obtained later. The time derivative of this expression evaluated at $t = t_0$ gives the initial condition for the radial velocity of the shock, $v_{r, \, 0}=\frac{(1-b_1)r_0}{3t_0}$. Likewise, the initial conditions for the axial position and velocity of the shock are, respectively, $d_0 = R_{c,0}\,r_0$ and $v_{d, \, 0} = R_{c,0}\,v_{r, \, 0}$ (where $R_{c,0}$ is a parameter to be fixed), and for the cocoon pressure $p_{c, \, 0}=0.4\,\frac{L_0\,t_0}{\pi\, R_{c,0} \, r_0^3}$.
Finally, $r_0$ is derived by applying the jump condition from the following expression:
\be
v_{r, \, 0}^2=\frac{(\Gamma+1) p_{c, \, 0}}{2 \rho\left(\sqrt{r_0^2+(d_0/2))^2}\right)}+\frac{(\Gamma-1)}{2 \Gamma}c_{s, \, a}^2.
\ee

Once the initial conditions are fixed, we solve the evolution differential equation \eqref{edo_r2} numerically using Mathematica, following these steps:
In the first phase, we impose $p_c = p_{c,0}(t/t_0)^{b_1}$ and apply the initial conditions described above, evolving the system until a time $t = t_1$ such that $\sqrt{r(t_1)^2 + (d(t_1)/2)^2} = l_{\rm crit,1}$ (where $l_{c,i}$ is the $i$-th position in the vector $l_{\rm crit}$ defined above; $a_g$ in this case). In the next phase, we replace $b_1$ with $b_2$ and $p_{c,0}$ with the value that ensures continuity of $p_c$. We impose as the initial condition the value of $r(t_1)$ obtained in the first phase and evolve the system until a time $t_2$ such that $\sqrt{r(t_2)^2 + (d(t_2)/2)^2} = l_{\rm crit,2}$. This process is repeated successively.

At this point, let us note that the pressure is non differentiable at the critical times $t_i$ ($i=1,... 4)$  due to the changes of the exponents $b_i$. So, from \eqref{marina}, $d(t)$ becomes also non differentiable, and thus the axial velocity $\dot d(t)$ becomes discontinuous. To avoid this problem, we make use of a mathematical operator equivalent to the piecewise function but smooth at the critical points. It is constructed as follows
\be\label{pc}
p_c(t)=p_{c,1}(t)^{\frac{1}{1+(t/t_1)^\eta}}\prod_{i=2}^4 \left(p_{c,i}(t)^{\frac{1}{1+(t_{i-1}/t)^\eta}\frac{1}{1+(t/t_i)^\eta }}\right)p_{c,5}(t)^{\frac{1}{1+(t_4 /t)^\eta}},
\ee
where
\bea
\label{pc1}
&&p_{c,1}(t)=p_{c0}\left(\frac{t}{t_0}\right)^{b_1},\\
\label{pci}
&&p_{c,i}(t)=p_{c,i-1}(t_{i-1})\left(\frac{t}{t_{i-1}}\right)^{b_i}.
\eea
for $i>1$. The coefficient $\eta$ regulates the smoothness. The larger it is, the better the function fits the original one, but the more abrupt the changes are. We choose $\eta=5$, which gives smooth enough results, and not very different from the ones obtained with a piecewise function.

Introducing this function into the differential equations~\eqref{marina} and \eqref{edo_r2} we can solve them for all times, with continuous and derivable functions, allowing us to obtain the evolution of the axial and transversal shock positions, $d(t)$, $r(t)$, in a consistent way, and the corresponding advance velocities $v_d(t)=\dot d(t)$, $v_r(t)=\dot r(t)$.

Finally, the pressure at the hotspot is obtained by directly applying the jump conditions at the interaction site, for which we use the following expression
\be \label{ph}
p_h(t)=\frac{2\rho(d(t))}{\Gamma+1}\left(v_d(t)^2-\frac{\Gamma-1}{2\Gamma}c_{s,a}^2\right).
\ee
Although this pressure is derived as the post-shock ambient one, it is the same as that of the post-shock jet (hotspot) pressure, because pressure is constant across the contact discontinuity between the two shocks.

The ambient pressures at the shock midpoint and at $d(t)$ are given, respectively, by, 
\bea \label{pac}
&&p_{ac}(t)=\frac{1}{\Gamma}\rho\left(\sqrt{r(t)^2+(d(t)/2))^2}\right)c_{s,a}^2,\\
\label{pah}
&&p_{ah}(t)=\frac{1}{\Gamma}\rho(d(t))c_{s,a}^2.
\eea

In short, our dynamical model for the supersonic phase can be summarized as follows. In order to solve for the four variables that describe the evolution of the system ($r, d, p_c,$ and $p_h$) we solve the equations of the jump conditions at the cocoon-ambient medium and at the hotspot-ambient medium interfaces (Eqs. \eqref{edo_r} and \eqref{ph}, respectively) and impose two restrictions, Eqs. \eqref{marina} and \eqref{pc} (with the cocoon pressures in the different phases defined as in Eqs.~\eqref{pc1} and \eqref{pci}). This system of equations is equivalent to Eqs. (1)–(4) in \citet{2018MNRAS.475.2768H}, except that we change one of the restrictions. Instead of imposing a relation between the hotspot pressure and the cocoon volume, we impose a particular function for the evolution of $p_c$, based on the results of previous numerical simulations.

\subsection{Transonic phase}

The equations and method described in the previous section are used as long as the post-shock pressures are significantly larger than the ambient pressure. However, note that when $p_c$ and $p_h$ approach $p_{ac}$ and $p_{ah}$, respectively, the radial and axial velocities must approach the ambient sound speed, consistent with equations \eqref{edo_r} and \eqref{ph}, and the subsequent evolution should take this into account. Furthermore, although the hotspot pressure, $p_h$, is initially larger than the cocoon pressure, $p_c$, there are cases (as in the case of rapidly expanding or low-power jets) in which $p_h$ falls faster than $p_c$, and they become equal along the course of evolution. Beyond this point, the model should account for the fact that the hotspot has disappeared and the shocked region should evolve as dictated by the cocoon pressure.

According to the arguments given in the previous paragraph, we should adapt our model to the following two situations:

\paragraph{The bow-shock becomes transonic.} For the case that $p_c$ reaches $p_{ac}$ before $p_h$ and $p_c$ become equal, we proceed as follows: 

\begin{itemize}

\item We define $t=t_{ac}$ as the time when $p_c=p_{ac}$ (or $v_r=c_{s,a}$). For $t>t_{ac}$, we impose that $v_r(t)=c_{s,a}$ and $r(t)=r_{ac}+c_{s,a}(t-t_{ac})$, where $r_{ac}$ is the value that $r$ had reached at $t_{ac}$. 
\item The relation \eqref{marina} is no longer valid, as it relies on the (supersonic) expansion of the cocoon being mediated by a strong shock. Instead, considering a mildly relativistic jet down to the hotspot, we can estimate the hotspot pressure as $p_h\sim\frac{L_0}{c\pi r_h^2}$, where $r_h$ is the jet radius at the hotspot. Assuming conical expansion, $p_h\propto d(t)^{-2}$. The proportionality constant of the relation is obtained by imposing continuity at $t_{ac}$. 

\item With this relation and the jump conditions between the hotspot and the ambient, we obtain the differential equation for $d(t)$ at $t>t_{ac}$. The initial conditions for the phase are set by imposing continuity of $d(t)$ at $t_{ac}$.
\be\label{edo_d}
\dot d(t)^2=\frac{(\Gamma+1) p_h(d(t))}{2 \rho(d(t))}+\frac{(\Gamma-1)}{2 \Gamma}c_{s,a}^2.
\ee
Since equations~\eqref{ph} and \eqref{edo_d} coincide at $t_{ac}$, the continuity of $v_d(t)=\dot d(t)$ at $t=t_{ac}$ is ensured.

\item Subsequently, we calculate the value of $t$ (which we define as $t_{ah}$) for which the hotspot reaches equilibrium with the ambient medium. From that moment, we impose $v_d(t)=c_{s,a}$ and $d(t)=d_{ah}+c_{s,a}(t-t_{ah})$, where $d_{ah}$ is the value $d(t_{ah})$. Finally, $p_{ac}$ and $p_{ah}$ evolve as dictated by Eqs.~\eqref{pac} and \eqref{pah}, and we impose $p_c=p_{ac}$ for $t>t_{ac}$, and $p_h=p_{ah}$ for $t>t_{ah}$.

\end{itemize}

\paragraph{The hotspot disappears.} The other possible option is that the cocoon and hotspot pressures become equal before reaching equilibrium with the ambient medium. In this case, we proceed as follows. The instant at which $p_c = p_h$ is expressed as $t_{ch}$. We then impose $p_h = p_c$ for $t > t_{ch}$, with $p_c$ given by the expression \eqref{pc}. This is consistent with the rapid homogenization of pressure in the cocoon. 

\begin{itemize}

\item First, we obtain $d(t)$ by applying the differential equation \eqref{edo_d}, but replacing $p_h(d(t))$ with $p_c(t)$. 

\item Subsequently, we obtain $r(t)$ by solving Eq.~\eqref{edo_r2}, but introducing the new function of $d(t)$ in the square root, instead of that derived from Eq.~\eqref{marina}. 

\item In both cases, we set initial conditions that ensure the continuity of $r(t)$ and $d(t)$ at $t_{ch}$. 

\item Then, as in the previous case, we calculate the values of time at which both velocity components reach the local sound speed. From those instants we impose $v_r(t) = c_{s,a}$ and $v_d(t) = c_{s,a}$, respectively, 
leading to a linear growth of both $r$ and $d$ with time.

\item Finally, following a similar procedure, we obtain $p_{ac}$ and $p_{ah}$ from Eqs.~\eqref{pac} and \eqref{pah}, and impose that $p_c$ and $p_h$ are equal to them (respectively), once the equilibrium with the ambient medium in each direction is reached.

\end{itemize}

\section{Evolution model: Radio luminosity}
\label{Rlum}

The radiative part of the model mainly follows \citet{1997MNRAS.292..723K} (hereafter referred to as KDA), but with three significant variations. First, in KDA the time dependence of the lobe pressure is expressed as a power law, whereas in our case it is a numerical function, which requires certain modifications to the equations as we describe below. Second, in our model, we consider the presence of thermal matter in the jet, and its proportion to be dependent on time/distance, and on jet power. Finally, KDA focuses on high power sources, and therefore does not consider the possibility that the shock may have vanished at the time of observation, that is, the hotspot and cocoon pressures may have already fallen to the ambient pressure. In our model, we propose a way to estimate the luminosity emitted by the material that the jet continues to inject into its lobes/plumes after the shock has disappeared.

Assuming that the electrons emit only at their critical frequency, the cocoon/lobe luminosity due to synchrotron radiation per unit of frequency is given by the following expression
\be \label{lum}
L_\nu (t)=\frac{2\sigma_{\mathrm{T}}c}{3\nu} \,u_B(t) \,\gamma(t)^3 N(t),
\ee
where $\sigma_T$ is the Thomson cross-section, $c$ is the speed of light, $u_B$ is the energy density of the magnetic field, $\gamma$ is the Lorentz factor of the relativistic electrons emitting at frequency $\nu$, and $N$ is the total number of emitting particles with the corresponding Lorentz factor. The time dependence of the magnitudes in \eqref{lum} comes from the dependence on the pressure of the cocoon. 

Synchrotron radiation theory \citep[e.g.,][]{1979rpa..book.....R} tells us that most of the radiation at frequency $\nu$ comes from electrons with a Lorentz factor
\be \label{gamma}
\gamma(t) =\sqrt{\frac{2 \pi m_{\mathrm{e}} \nu}{e B(t)}},
\ee
where $e$ and $m_e$ are the electron charge and mass, and $B$ is the magnetic field accelerating them. This, in turn, is related to the magnetic field energy by $u_B(t)=B(t)^2/(2\mu_0)$ where $\mu_0$ is the vacuum permeability.

In order to estimate the magnetic energy density, we proceed analogously to the KDA model. The total energy of the cocoon is the result of the sum of its main components: the magnetic field energy density, $u_B$, that of the emitting relativistic electrons/positrons, $u_e$, and that of the thermal matter (protons and non-relativistic electrons), $u_T$. In our model, we consider that the relativistic electrons and the magnetic field are in equipartition, i.e., $u_e(t) = u_B(t)$. On the other hand, we define the function $k(t) = u_T(t)/u_B(t)$, the ratio of thermal to magnetic (or non-thermal particle) energy densities. The total energy is then related to the cocoon pressure by $p_c = (\Gamma - 1)(u_B + u_e + u_T)$, where $\Gamma$ is an average adiabatic index of the cocoon that would depend on the relative weight of the three components. Combining these relations, we obtain the dependence of the energy densities on the cocoon pressure
\be
\label{eq:ub}
u_B(t)=u_e(t)=\frac{p_c(t)}{(\Gamma-1)(k(t)+2)}.
\ee

Regarding the fraction of thermal matter in the cocoon, $k(t)$, we assume that the jet entrains thermal matter as it expands within the galaxy. We approximate the jet mass-load by assuming that $k(t)$ grows linearly until the time it needs to reach three times the galactic core radius, at $t_{3a}$, i.e. $d(t_{3a}) = 3a$. This is a simple approach to account for mass-load by stellar winds and gas, which is expected to occur mainly within the central region of the host galaxy \citep[see, e.g.,][]{2014MNRAS.437.3405L,2020MNRAS.494L..22P}. From this moment onward, we assume that $k(t)$ remains constant. As for the limiting values of $k$, we consider that mass-loading is relatively more relevant in the case of low-power jets, as shown by, e.g., \cite{2014MNRAS.441.1488P,2021MNRAS.500.1512A}. We propose the following expression
\be \label{eq:load}
k(t)= \begin{cases} \displaystyle{c_k\frac{t-t_0}{\sqrt{L_0}}}, & \text { if } t \leq t_{3a} \\ \displaystyle{c_k\frac{t_{3a}-t_0}{\sqrt{L_0}}}, & \text { if } t>t_{3a} \end{cases}.
\ee
where $c_k$ is a constant to be defined. We have found that, for $c_k=10^6 \text{\,W}^{1/2}\text{\,s}^{-1}$, this expression gives terminal values of $k$ between 0.1 and 10 for powers greater than $10^{37} \text{W}$, and between 10 and 1000 for powers between $10^{35} \text{W}$ and $10^{37} \text{W}$. This gives the expected dominance of the thermal component in low-power, decelerated jets, in contrast to more powerful jets \citep{2014MNRAS.441.1488P,2021MNRAS.500.1512A,2023MNRAS.523.3583P}, as supported by previous observational estimates of particle content in radio lobes \citep{2003MNRAS.346.1041C,2008MNRAS.386.1709C,2008ApJ...686..859B}.

As far as the number of emitting particles is concerned, we need to estimate how many of those injected in the lobe have the appropriate Lorentz factor $\gamma(t)$. To this aim, we apply the approach described in KDA.

Following that work, we consider an instant $t_i$ (smaller than the observing time $t$) as the time when a volume element, $\delta \tilde{V}(t_i)$, is injected into the lobe, loaded with fresh particles, during an infinitesimal time interval $\delta t_i$. This volume element expands adiabatically in the lobe, reaching a value $\delta V(t_i, t)$ at the observing time $t$. On the other hand, taking $n(t_i, t)$ as the density of particles injected at $t_i$ that have a Lorentz factor $\gamma(t)$ at time $t$, and summing over all the volume elements that have been injected along the history of the source for times smaller than $t$, we obtain the total number of particles in the cocoon that have the appropriate Lorentz factor at the observing time, 

\be\label{N}
N_c(t)=\int_{t_0}^t n(t_i,t)\frac{\delta V(t_i,t)}{\delta t_i}dt_i.
\ee

The volume element $\delta \tilde{V}(t_i)$ can be obtained by assuming that it expands adiabatically during the interval $\delta t_i$, with pressure changing from the hotspot value, $p_h(t_i)$, to that of the cocoon, $p_c(t_i)$. One then obtains 
\be
\delta \tilde{V}(t_i)=\frac{\left(\Gamma-1\right) L_0}{p_{c}(t_{i})}\left(\frac{p_h(t_i)}{p_c(t_i)}\right)^{(1-\Gamma) / \Gamma} \delta t_{i} .
\ee
Subsequently, adiabatic expansion in the cocoon from $t_i$ to $t$ results in $p_c(t_i) \,\delta \tilde{V}(t_i)^\Gamma = p_c(t) \,\delta V(t_i, t)^\Gamma$, and therefore
\be
\frac{\delta V(t_i,t)}{\delta t_i}=\frac{\left(\Gamma-1\right) L_0}{p_{c}(t_{i})}\left(\frac{p_h(t_i)}{p_c(t_i)}\right)^{(1-\Gamma) / \Gamma} \left(\frac{p_c(t)}{p_c(t_i)}\right)^{-1/\Gamma},
\ee
which can be simplified as
\be\label{V}
\frac{\delta V(t_i,t)}{\delta t_i}=(\Gamma-1) L_0 p_c(t)^{-\frac{1}{\Gamma}}p_h(t_i)^{\frac{1}{\Gamma}-1}
\ee
(compare with Eq.~(13) in KDA).
 
The energy distribution of the particles at the moment of injection, $\tilde n$, is given by the power law
\be\label{dist_n}
\tilde{n} = n_0 \tilde\gamma^{-p},
\ee
where $n_0$ is a normalization constant, $p$ is a positive coefficient and $\tilde \gamma$ is the Lorentz factor of the particles at the moment of the injection. Energy losses cause a drop of the electron Lorentz factors, so, necessarily $\tilde \gamma>\gamma(t)$. 

The normalization number density of the emitting electrons, $n_0$, can be obtained from the energy density of the electrons $u_e(t)$, which is derived directly from the lobe pressure via the equipartition assumption. The particle kinetic energy of a relativistic electron is $m_e c^2 (\tilde\gamma - 1)$, and therefore the energy density of the electrons with Lorentz factor $\tilde\gamma$ at the moment of injection will be $m_e c^2 (\tilde\gamma - 1) \tilde{n}$. By integrating this expression over all possible energies, we can  equate the result of this integral to $u_e(t_i)$ and solve for $n_0$:
\be
n_0(t_i)=\frac{u_e(t_i)}{m_e c^2}\left(\int_{1}^{\infty}\left(\tilde\gamma-1\right) \tilde\gamma^{-p} \mathrm{~d} \tilde\gamma\right)^{-1}.
\ee
The spectral distribution of radio galaxies indicates $2 < p < 3$ \citep{10.1093/mnras/225.1.1}. For these values the integral is convergent, leading to
\be\label{n0}
n_0(t_i)=\frac{u_e(t_i)}{m_e c^2}(p-2)(p-1).
\ee

The particle density in the volume element is also affected by the adiabatic expansion. Hence, the particle density at time $t$ in an expanding volume $V$ is given by
\be\label{n}
n(t_i,t) =n_0(t_i) \frac{\tilde\gamma(t_i,t)^{2-p}}{\gamma(t)^2}\left(\frac{V(t)}
{V(t_i)}\right)^{-4/3}
\ee
(compare with Eq.~(9) in KDA). Applying the relation between pressure and volume for an adiabatic expansion ($V \propto p^{-1/\Gamma}$) we finally obtain
\be\label{n}
n(t_i,t) =n_0(t_i) \frac{\tilde\gamma(t_i,t)^{2-p}}{\gamma(t)^2}\left(\frac{p_c(t)}{p_c(t_i)}\right)^{4/(3\Gamma)}.
\ee
As mentioned, $\tilde\gamma(t_i,t)$ is the Lorentz factor that the particles must have had at $t_i$ to have $\gamma(t)$ at $t$. 

Synchrotron and Compton (due to interaction with the cosmic microwave background, CMB) losses imply a rate of change of the particles' Lorentz factor $\displaystyle{\frac{d\gamma}{dt}}$ given by $\displaystyle{-\frac{4\sigma_{\mathrm{T}}}{3m_e c} \gamma^2 \left(u_B + u_{C}\right)}$, where $u_C$ is the energy density of the CMB radiation, $u_C \approx 4.17 \times 10^{-14} \, (z+1)^4 \, \text{J m}^{-3}$. Regarding adiabatic losses, the loss rate for particles confined in an expanding sphere (corresponding to the volume element) is given by $\displaystyle\frac{d\gamma}{dt}=-\frac{\gamma}{r}\frac{dr}{dt}$, $r$ being the radius of the sphere \citep[see, e.g., equation 16.16 in][]{Longair_2011}. In addition, adiabatic expansion implies $V \propto p^{-1/\Gamma}$, where $V$ is the volume of the sphere representing the volume element ($V \propto r^3$), $p$ is the gas pressure and $\Gamma$ is the adiabatic index. Therefore, $r \propto p^{-1/(3\Gamma)}$ and the energy loss rate can be expressed as $\displaystyle\frac{d\gamma}{dt} = \frac{\gamma}{3\Gamma p} \frac{dp}{dt}$.

Finally, combining all those terms, we obtain the differential equation for the variation of the electrons' energy within the cocoon
\be \label{edo_gamma}
\frac{\mathrm{d} \gamma}{\mathrm{d} t}=\frac{\gamma}{3\Gamma p_c} \frac{\mathrm{d}p_c}{\mathrm{d}t}-\frac{4}{3} \frac{\sigma_T}{m_{e} c} \gamma^2\left(u_B+u_{C}\right) .
\ee
This expression recovers the one in KDA for $p_c\sim t^{-a_1 \Gamma}$. By imposing the initial condition that a particle's Lorentz factor is $\gamma(t)$ (given by \eqref{gamma}) at the observing time $t$, we can solve numerically the equation and derive the Lorentz factor that the particles must have had at $t_i$, $\tilde\gamma(t_i, t)$. It is noteworthy that, for a given $t$, $\tilde\gamma$ tends to infinity at a value $t_i=t_{min}<t$. This value must then be used as the lower limit of the time integral~\eqref{N} (as long as it is larger than $t_0$), since particles injected before this instant cannot cool down to $\gamma$ at time $t$, and therefore do not contribute to the observed luminosity.

Finally, introducing \eqref{V} and \eqref{n} in \eqref{N}, and then in \eqref{lum}, we obtain the following expression for the luminosity per unit frequency at time $t$ 
\be\label{lum_lobe}
L_{\nu, \, c}(t)=\frac{4 L_0\sigma_{\mathrm{T}}c\gamma(t)p_c(t)^{1+\frac{1}{3\Gamma}}}{3 \nu(k(t)+2)}\int_{\tilde t_{min}(t)}^{\,t}n_0(t_i) \tilde\gamma(t_i,t)^{2-p} \frac{p_h(t_i)^{-1+\frac{1}{\Gamma}}}{p_c(t_i)^{\frac{4}{3\Gamma}}}dt_i,
\ee
where $\tilde t_{min}(t)=\mathrm{max}(t_0,t_{\mathrm{min}}(t))$. Note that we have multiplied the original expression \eqref{lum} by a factor 2 in order to account for the emitting lobes of both jets, assuming symmetric injection at global scales. This integration takes some seconds to be solved numerically in Mathematica. 

This procedure also allows us to calculate the  volume of the  emitting material in the lobe at frequency $\nu$, at time $t$, by integrating the expression \eqref{V},
\be\label{vol_total}
V_c(t)=2(\Gamma-1) L_0 p_c(t)^{-\frac{1}{\Gamma}}\int_{\tilde t_{min}(t)}^t p_h(t_i)^{\frac{1}{\Gamma}-1} dt_i,
\ee
where, again, we use a factor $2$ to include the two lobes. Moreover, the model also gives the time evolution of the spectral index by considering two different frequencies $\nu$ and $\nu'$ and taking into account that $L_\nu/L_{\nu'}=(\nu/\nu')^{-\alpha_{\nu,\nu'}}$. Then
\be\label{alpha}
\alpha_{\nu,\nu'}(t)=-\frac{\log(L_\nu(t)/L_{\nu'}(t))}{\log(\nu/\nu')}.
\ee
If we did not consider radiative losses this expression would reduce to $\alpha_{\nu,\nu'}(t)=(p-1)/2$.

Along this whole section we have assumed that there exists a hotspot injecting particles into a cocoon. However, as we saw in the previous section, the dynamical part of this model also considers the phase in which the pressures of the hotspot and the cocoon equalize with that of the ambient medium—that is, the shock disappears, and the notion of "hotspot" and "cocoon" is lost. Nevertheless, as long as the jet remains active, it will continue injecting matter into the environment, creating plumes of radiating material. This suggests an idea to estimate the luminosity in this phase, by making use of expression \eqref{lum_lobe}. 

In these situations, we consider the jet itself as the injector, taking into account that particle acceleration occurs along the dissipation/deceleration region \citep{2014MNRAS.437.3405L}. The particles are then injected into the lobes, considered to start where the jet becomes transonic. Although the dissipation/deceleration region is probably non adiabatic, we use this simplified approach as an approximation to these sources in our model. Therefore, we apply the same luminosity expression as given before (Eq.~\ref{lum_lobe}), replacing $p_c$ by the ambient pressure at the jet head, i.e., $\rho_a(d(t)) c_{s,a}^2/\Gamma$, and $p_h$ by the jet pressure at the transonic region, which we assume to be the ambient pressure at half the jet length, i.e., $\rho_a(d(t)/2) c_{s,a}^2/\Gamma$. These changes are applied from the moment when the hotspot disappears as it becomes transonic, that is, for $t > t_{ad}$. This procedure is also applied to the calculation of the volume occupied by the emitting particles (Eq. \eqref{vol_total}).

\section{Parameter space}
\label{Param}

In this section, we present the parameter space used in our calculations. The evolution of the sources is followed up to $t = 1$ Gyr, which can be taken as a  conservative upper limit of the duration of the nuclear activity.
The following parameters have been fixed for all calculations: initial computing time $t_0 = 1000$~yr, particle energy distribution exponent $p = 2.14$ (as in KDA), and observation frequency $\nu = 150\,\text{MHz}$ (unless indicated otherwise).

The ISM and IGM are characterized by a fully ionized hydrogen plasma at temperature $T_a = 1.7\times 10^7K$. The sound speed is then computed using $c_{s, a}=\sqrt{\frac{\Gamma k_B T_a}{\mu m_H}}$, where $k_B$ is Boltzmann constant, $m_H$ is the atomic hydrogen mass, and $\mu = 0.5$ is the mass per particle in units of $m_H$. At the ambient temperature considered, the plasma is non-relativistic and, accordingly, $\Gamma = 5/3$.

The initial cocoon length-to-radius ratio, $R_{c,0 }= d_0/r_0$, is fixed depending on the jet power: $R_{c,0} = 4$ for $L_0\leq 10^{37}$ W and $R_{c,0} = 5$ for $L_0>10^{37}$ W. Concerning jet power, we explore a range from $10^{35} \text{W}$ to $10^{39} \text{W}$. Similarly, the host galaxy parameters vary within typical ranges: $a_{g,0} \in [0.5, 2] \text{\ kpc}$, $a_{c} \in [30, 60] \text{\ kpc}$, $\beta_g \in [-2, -1.6]$, $\beta_c$ $\in [-1.5, -1.1]$ and $\rho_{g,0} \in [0.1, 10] \ m_{\rm p}\ \text{cm}^{-3}$, where $m_{\rm p}$ is the proton mass, and $\rho_{c,0}$ is always taken to be $\rho_{g,0}/100$. In order to maintain the shape of the density profile with the 5 phases described in section \ref{shock_advance}, we choose combinations of values that verify $l_{gc}<a_c$. For simplicity, the viewing angle $\theta$ is fixed to $90^\circ$ throughout the entire work, and the redshift is set to $z = 0$, unless otherwise specified (note that for this value $a_g=a_{g,0}$).

\subsection{Lobe pressure evolution exponents}
\label{ss:lppe}

The exponents governing the time evolution of the cocoon pressure are also free parameters in our model. However, their range of variation can be constrained by assuming reasonable limits for the evolution of the axial and radial expansion speeds of the cocoon. In the following discussion, we assume that the cocoon is highly overpressured and its sideways expansion is correspondingly highly supersonic. According to this, the last term on the right-hand side of Eq.~\eqref{edo_r2} (for the cocoon expansion speed) is neglected when estimating the exponents of the power laws. 

For the initial expansion phase we have $p_c \propto t^{\ b_1}$. During this phase, the ambient density remains approximately constant, and we have $\dot r \propto p_c^{\ 1/2} \propto t^{\ b_1/2}$, and correspondingly $r \propto t^{\ b_1/2 +1}$. The axial expansion of the cocoon follows from Eq.~\eqref{marina}, $d \propto t \ p_c^{\ -1} r^{-2} \propto t^{-2b_1 - 1}$. The values of $b_1$ can be constrained by forcing that, on the one hand, the velocity of axial shock does not increase with time during this phase (and hence $b_1 > -1$), and, on the other, it falls at the same rate or slower than the speed of radial expansion (leading to $b_1 \leq -0.8$). In the end, we have chosen $b_1 \in [-0.9, -0.8]$. For the same reasons, we take the same interval for $b_5$, i.e., for the expansion phase through the homogeneous intergalactic medium, where the density is also constant.

In the second phase, the cocoon expands against an ambient medium in which density and pressure fall as $l^{\ \beta_g}$. During this phase, $p_c \propto t^{\ b_2}$ with the only restriction that both the axial and radial expansion speeds should evolve with time with powers between $0$ and $-1$ (corresponding to constant or decelerating expansion speeds). In establishing the range of values for $b_2$ that satisfy the previous conditions, we also consider that in the estimate of the radial expansion speed, the quantity $\sqrt{r^2 + (d/2)^2} $ must grow in time with an intermediate power between that of $ r $ and $ d $. If we assume, on the one hand, that $r$ increases faster than $d$ and that $\sqrt{r^2 + (d/2)^2}$ grows as $r$, we obtain the interval $ b_2 \in [(-4+\beta_g) / (5 + \beta_g), \,\beta_g]$ (for $\beta_g\in[-2,-1]$).  On the other hand, if we assume that $d$ grows faster than $r$ and that $\sqrt{r^2 + (d/2)^2}$ grows as $ d $, we obtain $ b_2 \in [\beta_g / 2 - 1, \,(-4+\beta_g) / (5 + \beta_g)] $. Therefore, the resulting interval that contains both possibilities is $ b_2 \in [\beta_g / 2 - 1, \beta_g] $.
We have verified that this condition results in exponents for the time evolution of the cocoon pressure which are in agreement with those obtained from numerical simulations \citep[e.g.,][]{2011ApJ...743...42P,2019MNRAS.482.3718P}.

For cases with a phase in which $ p_h = p_c $, i.e., for which the hotspot disappears, we have that $ r $ and $ d $ grow at the same rate. In this case, we obtain $ b_2 \in [-2, \beta_g] $. Note that the interval derived for the previous phase ($[\beta_g / 2 - 1, \,\beta_g]$) is contained in this new interval hence, for simplicity, we use the same exponent through both phases within the more restrictive interval.

We extend this procedure to the fourth phase, where $\rho\sim l^{\beta_c}$ and $p_c\sim t^{b_4}$. Analogously, in this phase we restrict the coefficient to the interval $ b_4 \in [\beta_c / 2 - 1, \,\beta_c] $. 

Regarding the third phase, where the cluster's density starts to dominate over the galaxy one, we do not have a given value for the exponent of the density. We will fit this parameter, which we will call $\beta_{gc}$ and, for simplicity, we will take the intermediate value for $b_3$, i.e., $b_3=(3\beta_{gc} - 2)/4$.

Let us remark that these intervals for the pressure exponents are only applied during the supersonic phase. When the cocoon reaches equilibrium with the ambient medium we impose $p_c$ to evolve with the ambient pressure, as explained above.

\section{Results} \label{results}

\subsection{The role of jet power} \label{sec:power}

In this section we present our study of the role of jet power on the evolution of radio galaxies, according to the dynamical model. We take the central values of the parameter space distributions for the characterization of the host galaxies. With this, we try to define an average galaxy. Therefore, we assume $a_{g} = 1.25$~kpc, $\beta_g = -1.8$, $\rho_{g,0} = 1.67\times 10^{-21}\,{\rm kg/m^{3}}$ (equivalent to $1\,m_p\,{\rm cm^{-3}}$), $a_c = 45$~kpc, $\beta_c = -1.3$, and $\rho_{c,0} \simeq 1.67\times 10^{-23}\,{\rm kg/m^{3}}$ (equivalent to $0.01\,m_p\,{\rm cm^{-3}}$). We also fix the cocoon pressure exponents to the central values within the defined intervals, $b_1 = b_5 = -0.85$ along the regions with constant density, $b_2 = -1.85$ for the galactic profile, 
and $b_4 = -1.475$ for the group/cluster profile. $b_3$ is, as explained above, always fixed at $(3\beta_{gc}-2)/4$, where $\beta_{gc}$ is obtained by fitting the density profile in that phase. For these parameters we have $\beta_{gc}\approx-0.84$ and $b_3\approx-1.13$. 

\begin{figure}[htbp] 
\begin{center}
\includegraphics[width=90mm]{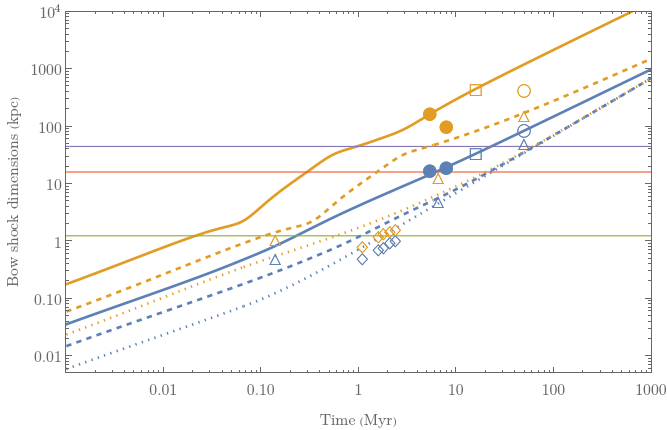}
\end{center}
\caption{Evolution of the shock radius $r(t)$ (blue) and head position $d(t)$ (orange) for different jet powers $L_0$: $10^{39}$ W (solid curves), $10^{37}$ W (dashed curves) and $10^{35}$ W (dotted curves). The horizontal lines, limiting the evolution phases 1 to 4, indicate the galactic core radius, $a_g$, (green), the position at which the galactic density reaches the cluster one, $l_{gc}$ (red), and the cluster core radius, $a_c$ (purple). 
The symbols indicate the values obtained from multidimensional relativistic (RHD) simulations. The empty symbols represent two-dimensional, axisymmetric numerical simulations: The rhombs stand for the $5\times10^{34}$~W simulations in \citet{2014MNRAS.441.1488P}, the triangles for the $10^{37}$~W simulations in \citet{2007MNRAS.382..526P}, \citet{2014MNRAS.441.1488P} and \citet{2014MNRAS.445.1462P}, and the circles and squares for the $10^{38}$ and $10^{39}$~W simulations, respectively, in \citet{2014MNRAS.445.1462P}. The solid symbols indicate the results obtained from three-dimensional simulations \citep{2019MNRAS.482.3718P,2022MNRAS.510.2084P,2023MNRAS.523.3583P}. As for the lines, blue symbols correspond to shock radii and orange ones to head positions.}
\label{size}
\end{figure}

Figure~\ref{size} shows the evolution of $d$ (the tip of the bow shock; orange lines) and $r$ (its mean radial coordinate; blue lines) with time for different jet powers. The figure shows the evolution of radio galaxies for three different jet powers, $10^{35}$, $10^{37}$, and $10^{39}$~W, in dotted, dashed, and solid lines, respectively. According to these results, for the parameters considered, sources with powers $10^{39}$~W would reach 1~Mpc at a time of the order of 100~Myr, whereas those with $10^{37}$~W would require long lifetimes ($~\sim 1$~Gyr) to reach megaparsec scales. 

The cocoon length-to-radius ratio, $R_c = d/r$, reaches values $\simeq 12$ for powerful jets ($L_j\geq 10^{38}\,{\rm W}$, within the first Myr of evolution) and $\simeq 11$ for intermediate-power ones ($L_j\simeq10^{37}\,{\rm W}$, after a few Myr). After this time, the former maintains values between 9 and 15, while the latter drops monotonously to a value of 2 at $t\,=\,1{\rm  Gyr}$. This agrees with numerical simulations of jets evolving in similar environments \citep[e.g.,][]{2011ApJ...743...42P,2014MNRAS.441.1488P,2014MNRAS.445.1462P,2019MNRAS.482.3718P}. Finally, as far as low power sources is concerned, the figure shows that they tend to approximately spherical expansion within 10 Myr, with similar values for axial and radial lengths \citep[in agreement with, e.g.,][]{2014MNRAS.441.1488P,2016A&A...596A..12M}. The almost parallel evolution of the $r$ and $d$ lines along phases 1 and 4, corresponding to an almost logarithmically linear growth of both quantities with time, is a consequence of the values chosen for the exponents $b_1$ and $b_4$. They are close to the values that correspond to self-similar evolution, which are $-4/5$ and $(-4+\beta_c)/(5+\beta_c)$  respectively. Indeed, we have verified that within the plausible combinations of $\beta_g$, $\beta_c$ and the exponents of the cocoon-pressure, some of them result in approximate self-similar expansions, whereas others show a time dependence of $R_c$.

In the figure, the symbols indicate the values obtained from multidimensional relativistic (RHD) simulations. Interestingly, the symbols fall within the region limited by the expected evolutionary tracks. The dispersion and shifts can be explained by differences in the properties of the host galaxies or jet head deceleration by mass loading in the low power jets \citep{2014MNRAS.441.1488P}. 

%
%

\begin{figure}[htbp]
\begin{center}
\includegraphics[width=90mm]{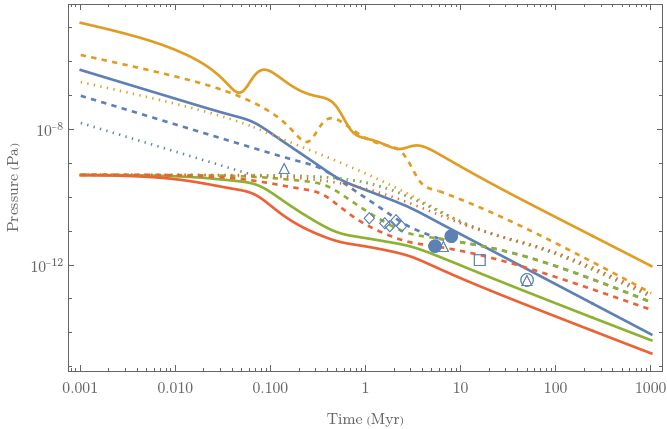}
\end{center}
\caption{Evolution of the hotspot pressure $p_h(t)$ (orange), cocoon pressure $p_c(t)$ (blue), ambient pressure at the position of the hotspot, $p_{ah}(t)$ (red), and of the cocoon midpoint, $p_{ac}(t)$ (green). We represent them for different jet's powers $L_0$: $10^{39}$~W (solid curves), $10^{37}$~W (dashed curves) and $10^{35}$~W (dotted curves). 
The points indicate cocoon pressures obtained from numerical simulations, and are classified as explained in the caption of Fig.~\ref{size}.}
\label{pressure}
\end{figure}

Figure~\ref{pressure} shows the evolution of the hotspot pressure (orange), cocoon pressure (blue), ambient pressure at the hotspot location (red), and ambient pressure at the radial reference point of the shock (green) versus time for different jet powers ($10^{39}$ W, solid curves, $10^{37}$ W, dashed curves, and $10^{35}$ W, dotted curves). The convergence between the blue and green dotted lines (low power jets) at $t\simeq 0.1$~Myr shows that pressure equilibrium has been reached between the lateral shock front and the local ambient medium. Comparing this with Fig.~\ref{size}, we see that this happens within the host galaxy, at $\leq 1$~kpc from the active nucleus. This means that the shocks must become transonic within this region. In the case of the orange and red dotted lines, indicating the hotspot and ambient pressure at its location, the convergence takes place before 10~Myr.

The convergence of the pressure evolution to the ambient values indicates an upper limit of the shock or hotspot lifetimes and the moment from which the radio galaxy expands at the speed of sound of the ambient medium. As in the case of the evolution of the radial and axial expansion of the cocoon (Fig.~\ref{size}), the fact that the lines of the hotspot pressure during phases 1 and 4 are almost parallel to those of the cocoon pressure is a consequence of choosing values for the pressure exponents close to those corresponding to self-similar evolution.   

%
%

\begin{figure}[htbp]
\begin{center}
\includegraphics[width=90mm]{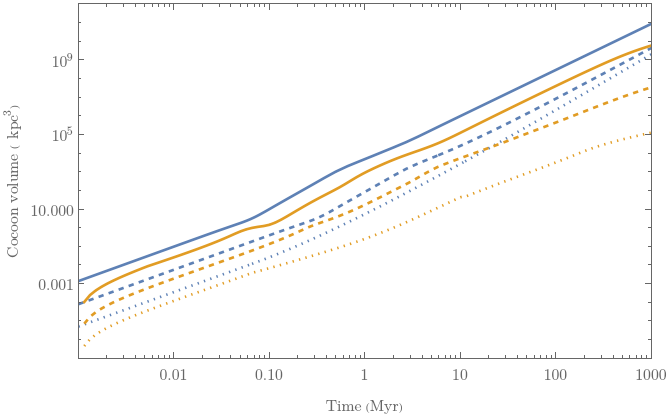}
\end{center}
\caption{Evolution of the cocoon's total volume (blue) and the emitting volume $V_c(t)$ (orange) for different jet powers $L_0$: $10^{39}$ W (solid curves), $10^{37}$ W (dashed curves) and $10^{35}$ W (dotted curves).}
\label{volume}
\end{figure}

Figure~\ref{volume} shows the evolution of the shocked ambient volume, i.e. $2 \pi r(t)^2 d(t)$ (blue lines), as a function of time, together with that of the emitting volume (the volume occupied by the particles injected into the cocoon that emit at $150$~MHz, as given by Eq.~\ref{vol_total}; orange lines), for the three selected powers. In agreement with simulations, the shocked ambient shell is relatively thin around the cocoon for powerful jets, whereas the difference between the volumes becomes large for weaker jets \citep[e.g.,][]{2014MNRAS.445.1462P}.

%
%

\begin{figure}[htbp]
\begin{center}
\includegraphics[width=90mm]{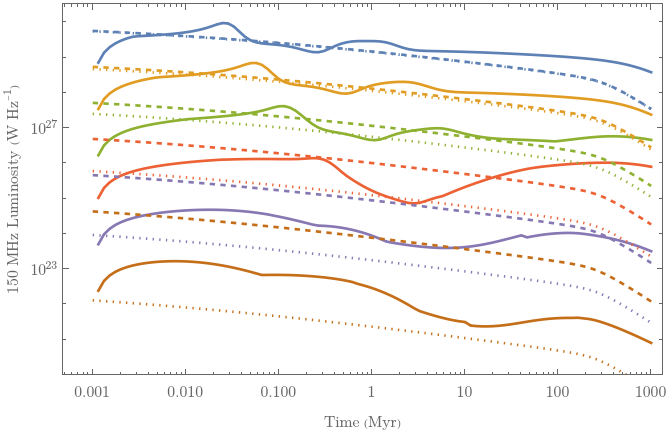}
\end{center}
\caption{Evolution of the luminosity at frequency $150$~MHz for different jet powers $L_0$: $10^{40}$ W (blue), $10^{39}$ W (orange), $10^{38}$ W (green), $10^{37}$ W (red), $10^{36}$ W (purple) and $10^{35}$ W (brown). We compare the results of our model (solid curves) with the KDA ones (dashed curves for $k=0$ and dotted curves for $k=c_k(t_{3a}-t_0)/\sqrt{L_0}$).}
\label{lumt}
\end{figure}

Figure~\ref{lumt} shows the evolution of the luminosity of radio galaxies at $150$~MHz for jet powers ranging from $10^{35}$ to $10^{40}$~W as derived by our model. The dashed and dotted lines in Fig.~\ref{lumt} show the luminosity evolution according to the original KDA model in two limiting mass load cases. The dashed lines show the evolution for purely non-thermal jets (coefficient $k$, the ratio of thermal to non-thermal particle energy densities, equal to zero), whereas the dotted ones incorporate the effects of a thermal component in the cocoon through the maximum value of $k$ in Eq.\eqref{eq:load}, $k=c_k(t_{3a}-t_0)/\sqrt{L_0}$.

The figure shows that the evolution of the total luminosity depends strongly on the jet power, and the trend depends on the ambient pressure/density profile. The initial evolution is dominated by a phase of rise \citep[e.g.][]{2000MNRAS.319..445S}, coincident with the propagation through the galactic core (compare with Fig.~\ref{size}). The duration of this phase is thus inversely correlated with the jet power. This phase is followed by a slow decreasing phase as the radio source evolves through the galactic pressure slope. Interestingly, we observe a similar evolution to that predicted by KDA for powerful jets as they develop along the galactic profile (phase 2 in our model). The transition to the group/cluster core plus a smoother slope after this core completely change the evolution. We also see how mass-load changes the radiative output mainly for low power jets, in contrast to powerful ones, which are basically unaffected. The two limiting cases of the KDA model considered give luminosities that enclose those obtained with our model for the largest part of the evolution, for powers $\leq 10^{37}\,{\rm W}$.

In the low-power curves (purple and red), we can observe a slight discontinuity between $10$ and $100$ Myr. This occurs because, at those times, the pressure of the hotspot reaches equilibrium with the ambient medium (see Figure \ref{pressure}), and from that moment onward we apply the change in the luminosity calculation described at the end of sec.~ \ref{Rlum} (we replace the cocoon with the remaining lobe/plume at the head of the jet). The jump is relatively smooth, which is due to the pressures of the regions under consideration being similar –as they essentially correspond to the ambient pressure at different locations–, immediately before and after the transition. This results in a smooth transition, as one would expect from a physically realistic evolution.

%
%

\begin{figure}[htbp]
\begin{center}
\includegraphics[width=90mm]{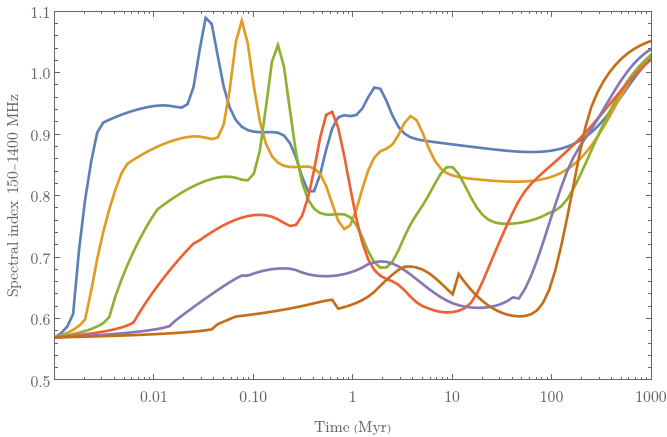}
\end{center}
\caption{Evolution of the spectral index between frequencies $150$~MHz and $1400$ MHz for different jet powers $L_0$: $10^{40}$ W (blue), $10^{39}$ W (orange), $10^{38}$ W (green), $10^{37}$ W (red), $10^{36}$ W (purple) and $10^{35}$ W (brown).}
\label{spectral}
\end{figure}

Figure~\ref{spectral} shows the evolution of the spectral index between frequencies $150$~MHz and $1400$ MHz, as computed from Eq.~\eqref{alpha}, for different jet powers. In all cases, there is an initial phase in which the spectrum steepens fast, associated with the expansion of the radio lobe through the galactic core (see Fig.~\ref{size}). From this point onward, the spectral indices undergo large variations (overall those associated to the high-power jets) of several tenths, with steep rises and falls. These variations are due to the fact that the rapid increases and decreases in luminosity occurring during the phase transitions do not take place simultaneously for the two frequencies, but are instead out of phase. Finally, after 100~Myr, the inverse Compton process starts dominating the losses and the spectral indices rapidly steepen and converge to $\alpha=p/2$ \citep[the mathematical proof of the convergence of the spectral index at late times in our model is given in the appendix \ref{assymptotic}; see also][]{1970ranp.book.....P}. This convergence will be relevant when discussing the evolution of high-redshift sources in a cosmological context, but it is not expected to be reached at low redshifts, but for extremely old radio galaxies.

%
%

\begin{figure}[htbp]
\begin{center}
\includegraphics[width=90mm]{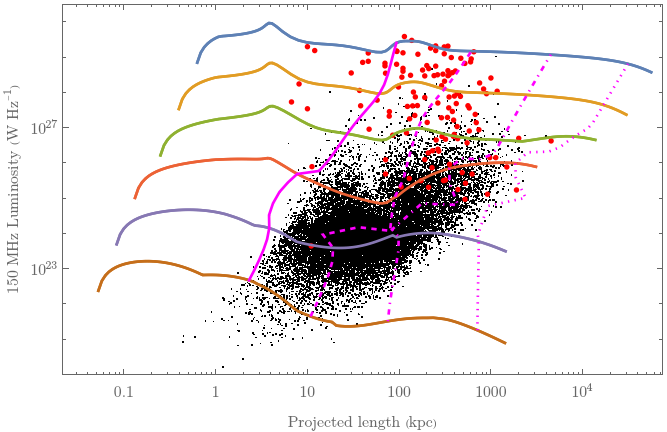}
\end{center}
\caption{Luminosity at frequency $150$~MHz as a function of the source axial size for different jet powers $L_0$: $10^{40}$ W (blue), $10^{39}$ W (orange), $10^{38}$ W (green), $10^{37}$ W (red), $10^{36}$ W (purple) and $10^{35}$ W (brown). LOFAR catalogue radio galaxies are represented by black points and the 3C catalogue sources by red points. The pink lines indicate isochrones for 0.5~Myr (solid), 5~Myr (dashed), 50~Myr (dot-dashed), and 500~Myr (dotted).}
\label{LumvsDobs}
\end{figure}

In order to compare our results with radio galaxies in the $P-D$ diagram \citep{1982IAUS...97...21B}, we represent the evolution of luminosity as a function of the projected length of the source in Fig.~\ref{LumvsDobs}. The projected length of a source is obtained by the expression $2d(t)\sin{\theta}$, where $\theta$ is the jet viewing angle ($90^\circ$ in our case). Evolution profiles are computed up to $t=1000$~Myr. We have added the values for the radio galaxies in the 3CR \citep{1983MNRAS.204..151L} and LoTSS catalogs \citep{2019A&A...622A..12H}. The pink lines indicate isochrones for 0.5~Myr (solid), 5~Myr (dashed), 50~Myr (dot-dashed), and 500~Myr (dotted). Although projection effects and/or changes in the characterization of the host galaxy would certainly produce a spreading in the location of the isochrones on this plot (see sec.~\ref{sec:host}), we would like to note that our model limits the ages of radio galaxies between $\sim 50$ and $\sim 500$~Myr, in agreement with other models \citep[e.g.][]{2015ApJ...806...59T}.  

%
%

\subsection{The host galaxy} \label{sec:host}

\begin{figure}[htbp]
\begin{center}
\includegraphics[width=90mm]{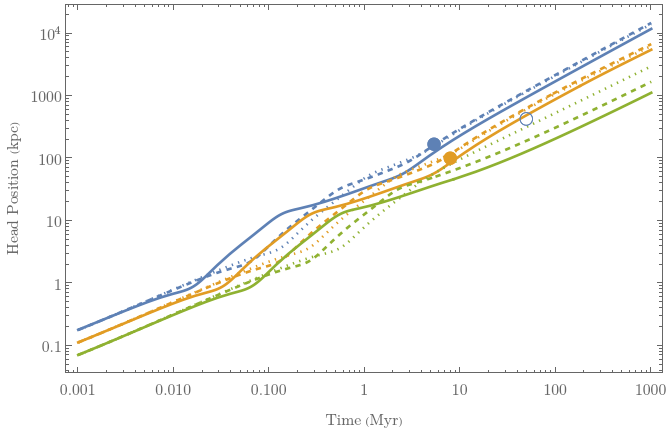}
\end{center}
\caption{Evolution of the radio galaxy head position, $d(t)$, with fixed jet power ($10^{38}$ W) for different host galaxy parameters: $\rho_{g,0}=0.1\,m{\rm _p/cm^3}$ (blue), $\rho_{g,0}=1\,m{\rm _p/cm^3}$ (orange), and $\rho_{g,0}=10\,m{\rm _p/cm^3}$ (green); in all cases $\rho_{c,0}=\rho_{g,0}/100$; $(a_{g},a_c)=(0.5,30)$~kpc (solid), $(a_{g},a_c)=(1.25,45)$~kpc (dashed) and $(a_{g},a_c)=(2,60)$~kpc (dotted). The points indicate results obtained from numerical simulations: the empty circle correspond to a 2D simulation, whereas solid circles represent 3D simulations. The colors indicate which line must be compared with each point (see more details in the text).}
\label{lengthrho0a}
\end{figure}

In this section we study the role of the properties of the hosts on the evolution of radio galaxies, by fixing jet power to $10^{38}$~W. Figure~\ref{lengthrho0a} shows the evolution of the radio galaxy axial size with time for different values of the galactic core densities ($\rho_{g,0} = 0.1, 1, 10 \,m_p\,{\rm cm^{-3}}$ represented by blue, orange and green lines, respectively, with $\rho_{c,0}=\rho_{g,0}/100$ in all cases), and core sizes ($\{a_{g},a_{c}\} = \{0.5, 30\},\, \{1.25, 45\}$, and $\{2, 60\}$ –all in kpc–, represented by solid, dashed and dotted lines, respectively). The corresponding masses for these hosts and their groups/clusters --calculated as the mass enclosed within $R_{200}$, the radius at which that the density is $200$ times the critical density-- are between $10^{9}$ and $10^{10}\,{\rm M_\odot}$ for the smallest densities (with $R_{200}\in [20,70]$ kpc), between $10^{12}$ and $10^{13}\,{\rm M_\odot}$ for the intermediate ones (with $R_{200}\in [250,550]$ kpc), and between $10^{14}$ and $10^{15}\,{\rm M_\odot}$ for the densest hosts (with $R_{200}\in [1500,3000]$ kpc). In this plot, we have kept the values of the ambient density and, consequently, lobe pressure evolution exponents as in the previous section ($\beta_g=-1.8$, $\beta_c=-1.3$, $b_1=b_5=-0.85$, $b_2=-1.85$ and $b_4=-1.475$). The figure shows that the smaller the values of the central densities the farther is the position reached by the jet's head at a given time (partially because of the influence of $\rho_{g,0}$ in the initial conditions). On the other hand, larger galaxy cores imply larger times for the jet to reach the second phase of evolution (with the jet's head propagating across the galactic atmosphere; first change of slope in each curve). Moreover, this phase (between the first and the second change of slope in each curve) is longer for larger cluster cores, which implies reaching a given distance at larger times.

The symbols show the position for the 2D jet simulation presented in \citet{2014MNRAS.445.1462P} (empty blue circle), the 3D simulations in \citet{2019MNRAS.482.3718P} (blue circle), and \citet{2022MNRAS.510.2084P} (orange circle). The three cases have been chosen because of their jet power, $10^{38}$~W. The simulated jets evolve through a galactic ambient medium defined by a double core with radii $a_0=1.2$~kpc for the galaxy and $a_1=52$~kpc for the group. For the blue symbols, the central density is $\rho_0=0.18\,m{\rm _p/cm^3}$, so they should lie close to the dashed blue line. We see that both points are indeed close to the dashed blue line, with the 3D simulation closer to the model prediction. Regarding the orange circle, we considered $\rho_0=0.72\,m{\rm _p/cm^3}$ in the corresponding 3D simulation, which should bring the evolution closer to the dashed, orange line. We see that again the model succeeds in giving a good approximation of the size of the radio galaxy.

%
%

\begin{figure}[htbp]
\begin{center}
\includegraphics[width=90mm]{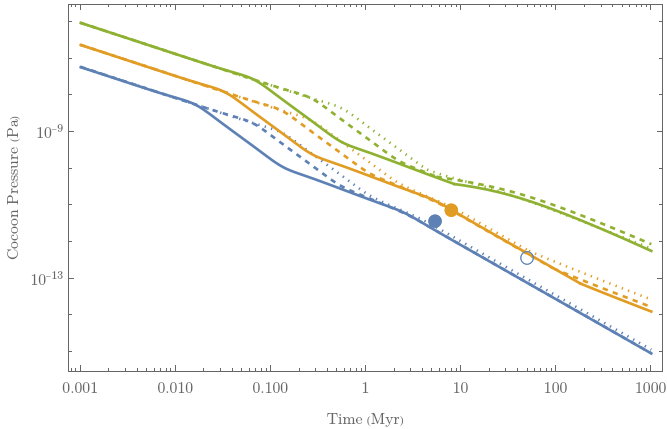}
\end{center}
\caption{Evolution of the cocoon's pressure $p_c(t)$ for a jet of $10^{38}$ W and different host galaxy parameters as described in Fig.~\ref{lengthrho0a}. The points indicate results obtained from numerical simulations, and are classified as explained in Fig.~\ref{lengthrho0a}.}
\label{pcrho0a}
\end{figure}

Figure~\ref{pcrho0a} shows the evolution of the pressure in the shocked region with time for the same jet power and different galactic profiles, as used in Fig.~\ref{lengthrho0a}. Comparing both figures we can easily verify the inverse proportionality between cocoon's pressure and head position. As expected, more diluted media result in smaller pressures at a given time. The cores, however, have limited influence on the pressure evolution, except for advancing or retarding the different phases. The change of slope of the green curves at $\sim10$ Myr and the orange curves at $\sim 100$ Myr is remarkable. It is caused by the cocoon pressure reaching equilibrium with the ambient pressure and evolving with it.
The points represent the numerical simulations mentioned above, and must be compared with the lines in the same way. We see that the model also succeeds in giving a reasonable approximation of the cocoon pressures mainly in the case of 3D simulations.

%
%

\begin{figure}[htbp]
\begin{center}
\includegraphics[width=90mm]{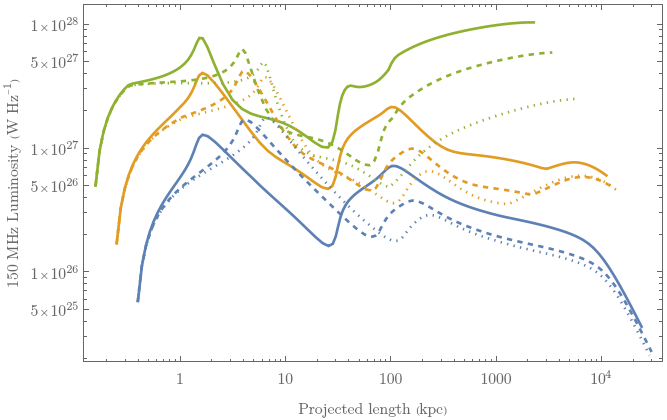}
\end{center}
\caption{$150$~MHz Luminosity vs projected length for a jet of $10^{38}$ W and different host galaxy parameters as described in Fig.~\ref{lengthrho0a}.}
\label{lumDrhoa}
\end{figure}

\begin{figure}[htbp]
\begin{center}
\includegraphics[width=90mm]{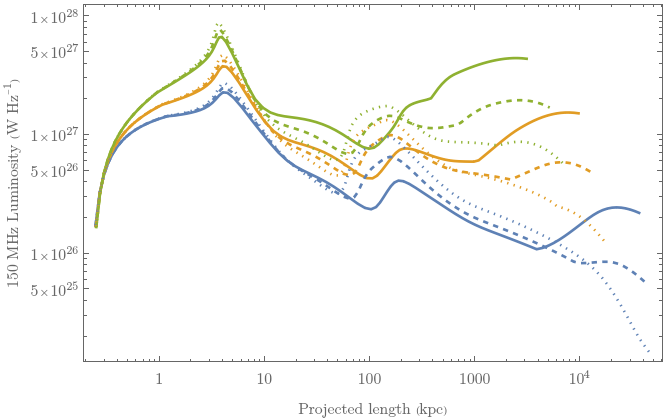}
\end{center}
\caption{$150$~MHz Luminosity vs projected length for different profiles of ambient density and cocoon pressure: $(\beta_g,\beta_c)=(-1.6,-1.1)$ (solid line), $(\beta_g,\beta_c)=(-1.8,-1.3)$ (dashed line) and $(\beta_g,\beta_c)=(-2,-1.5)$ (dotted line); $b_1=b_5=-0.9$, $b_2=\beta_g/2-1$ and $b_4=\beta_c/2-1$ (blue), $b_1=b_5=-0.85$, $b_2=(3\beta_g-2)/4$ and $b_4=(3\beta_c-2)/4$ (orange), $b_1=b_5=-0.8$, $b_2=\beta_g$ and $b_4=\beta_c$ (green).}
\label{lumDbetabb2}
\end{figure}

\begin{figure}[htbp]
\begin{center}
\includegraphics[width=90mm]{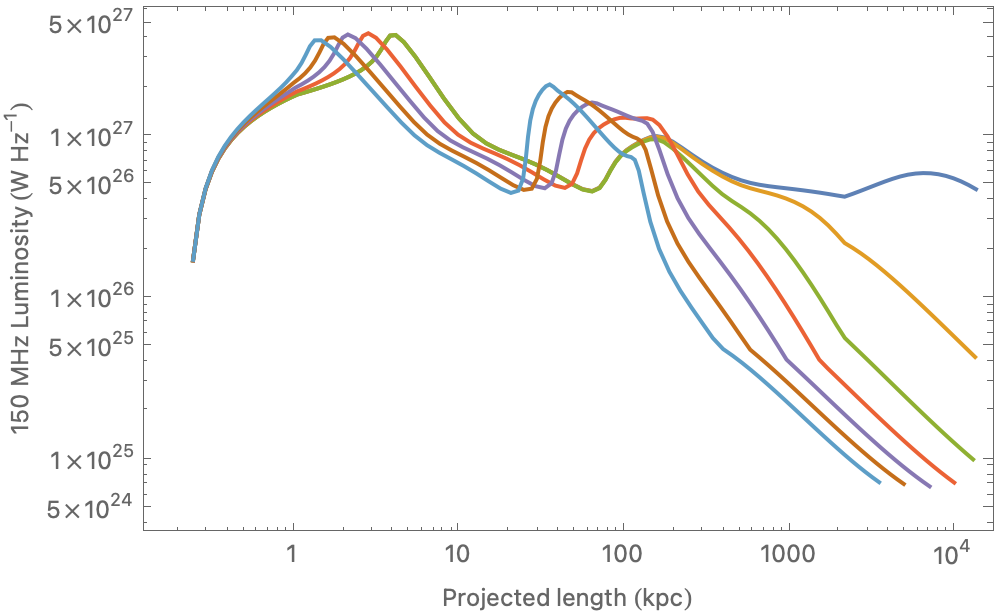}
\end{center}
\caption{$150$~MHz Luminosity vs projected length for different redshifts: $z=0$ (blue), $z=1$ (orange), $z=2$ (green), $z=3$ (red), $z=4$ (purple), $z=5$ (brown), $z=6$ (light blue).}
\label{lumDz}
\end{figure}

The next three figures show the evolution of lobe luminosity at 150~MHz with source length when varying different parameters: core size and central density (Fig.~\ref{lumDrhoa}), $\beta$-profiles and pressure time-evolution exponents (Fig.~\ref{lumDbetabb2}), and redshift (Fig.~\ref{lumDz}). Figure~\ref{lumDrhoa} shows the evolution of luminosity for a $10^{38}$~W radio galaxy evolving through the different hosts described above. Their behaviour is comparable to that observed in Fig.~\ref{LumvsDobs}, although this plot shows how sensitive the luminosity can be to the host properties. A comparison with Fig.~\ref{pcrho0a} reveals the expected correlation between radio luminosity and cocoon pressure. All models show an initial increase in luminosity, followed by a fast drop as the source expands out of the galactic core. This phase ends as the radio galaxy approaches the flatter gas distribution of the group/cluster core. The change is more abrupt for denser and/or more compact cores. The most extreme case corresponds to galactic and group/cluster core densities of 10 and 0.1~cm$^{-3}$, respectively, for which the luminosity rises by an order of magnitude from its minimum. At these high densities and the extended shallow density profiles of the group/cluster medium, the luminosity keeps growing between 100~kpc and 1~Mpc. In the other cases, the luminosity reaches a second maximum at distances $\geq 100$~kpc and then falls smoothly until Compton losses take over in more dilute media, or a new, minor growth occurs beyond 1~Mpc due to the entrance into the constant density region.

Figure~\ref{lumDbetabb2} shows the luminosity evolution of a $10^{38}$~W radio galaxy through the ambient medium with different $\beta$ profiles, also causing a change in pressure exponents, $b_i$. For these calculations, we fix the galactic core radius to the intermediate values for density and core sizes used in this section ($a_{g}\,=\,1.25$~kpc, $a_{c}\,=\,45$~kpc,  $\rho_{g,0}\,=\,1\,m{\rm _p/cm^3}$ and $\rho_{c,0}=\rho_{g,0}/100$). The solid lines track the evolution for $\beta_g = -1.6$ and $\beta_c = -1.1$; the dashed lines for $\beta_g = -1.8$ and $\beta_c = -1.3$, and the dotted ones for $\beta_g = -2$ and $\beta_c = -1.5$. The pressure evolution exponents along the constant density regions ($b_1$ and $b_5$) are always between $-0.9$ and $-0.8$. Along the density/pressure slopes ($b_2$ and $b_4$), they are taken between $\beta/2-1$ and $\beta$ (with $\beta = \beta_g, \beta_c$, for $b_2$, $b_4$, respectively; see sec.~\ref{ss:lppe}). The different colours of the lines indicate the values of the exponents considered: the most negative values (faster drops in pressure) are indicated by blue lines, the intermediate values by orange lines, and the highest values (slower drops in pressure) by green lines. For $b_3$ we always use the intermediate value of the interval, as derived from the $\beta$ fitted for the region.

During the initial phases, while the jets evolve through the galactic core and out of it, the differences between the models are driven by the values of $b_1$ and $b_2$. Beyond this inner region the density exponents play a major role. We see that the dashed and solid orange and green lines, with shallower drops in density and slow pressure evolution, produce significant increases in luminosity beyond 100~kpc, which persist through long distances. In contrast, steep ambient density profiles (dotted lines) and faster pressure evolution (blue lines), show a decreasing trend beyond the core/cluster cores. The rise of the solid and dashed lines in the last phase of the evolution is a consequence of the cocoon's pressure reaching equilibrium with the ambient medium. From this instant on, the pressure decreases more slowly, causing the change in slope. 
We therefore see that even for the same jet power and core densities, the luminosity evolution of the radio galaxy can be very different, depending on the ambient gas distribution.

Finally, Fig.~\ref{lumDz} shows the influence of redshift on the luminosity evolution of radio galaxies. The fiducial values are as before: $a_{g,0}\,=\,1.25$~kpc, $a_{c}\,=\,45$~kpc,  $\rho_{g,0}\,=\,1\,m{\rm _p/cm^3}$ and $\rho_{g,0}\,=\rho_{c,0}/100$, and the exponents $\beta$ and $b$ for density profile and cocoon pressure evolution are also fixed to their reference values. Then, we change $a_{g}$ (see Eq.~\ref{eq:core}) and $\rho_{\rm IGM}$ for different redshifts, as explained in sec.~\ref{sec:host1}. The figure shows that the smaller cores with increasing redshift displace the radio peak towards smaller scales, whereas inverse Compton losses become dominant earlier, so the radio luminosity is quenched at smaller scales.

%
%

%
%

%
%

\section{Discussion}
\label{Discussion}

\subsection{Comparison with other models}

In this section we compare our results to previous radio galaxy evolution models. Following seminal works by \cite{1974MNRAS.166..513S} and \cite{1982IAUS...97...21B}, a number of models have been published with apparently small, but relevant differences. \cite{1987MNRAS.226..531G} introduced jet opening angles, a transition to a constant density IGM and a redshift dependence of the ambient density ($\propto (1+z)^3$) to study the evolution of powerful sources. The authors use a core number density of $10^{-2}\,{\rm cm^{-3}}$, which changes to $7\times10^{-7}\,{\rm cm^{-3}}$ (for $z=0$) at the transition to the IGM at $\sim 200$~kpc. They predict radio galaxy ages of $10-100$~Myr for sizes $\geq 100$~kpc. Therefore, observed giant radio galaxies would require galactic activity to be sustained for $10^8-10^9$~yr. In \cite{1989MNRAS.239..173G}, the authors force a smaller jet opening angle than in the previous work, which, together with the low core density, allows radio galaxies to expand faster and reach $\sim 1$~Mpc in $10-100$~Myr in the case of powerful sources. This is closer to our results on the ages and sizes of large-scale radio galaxies, although the parameters that the authors use for the host galaxy ($\rho_0 = 10^{-2}m_p\,{\rm cm^{-3}}$ and $a_0 = 2$~kpc) are slightly different to ours.

The radio luminosity evolution derived from both our model and that of \cite{1997MNRAS.286..215K}, \cite{1997MNRAS.292..723K} is compared in Fig.~\ref{lumt}. In terms of radio galaxy dynamics, the main difference is that the KDA model is a self-similar model with a single $\beta$-profile. Early evolution is different between both evolutionary models because the KDA model does not include the expansion phase across the galactic core, making the radio luminosity fall monotonically in their case, as opposed to our model, which gives an initial increase. This agrees with the results provided by \citet{2000MNRAS.319..445S} or \citet{2002ApJ...568..639P}. Beyond this point, we observe that the evolution is similar for powerful jets as long as the jet evolves through the galactic medium, but already significantly different below $10^{37}$~W. The second $\beta$-profile completely changes the evolution between the two models, producing an increase in the luminosity in our case as compared to the continuous fall in theirs. Regarding the jet composition, it is important to note that in our case, the fraction of thermal to non-thermal energy density evolves in time as a consequence of baryon loading within the host galaxy, but it is fixed in the KDA models. To make the comparison between the models more reliable we have calculated the evolution according to the KDA model in two limits: zero mass-load (pure non-thermal composition) and maximum mass-load as given by the expression applied to our models (Eq.~\ref{eq:load}). We observe that the KDA lines typically lie above and below the luminosities predicted by our model, respectively. In the case of powerful jets, mass-load is dynamically insignificant, and the predicted luminosities are thus closer to each other.

\citet{1999AJ....117..677B} and \citet{2002A&A...391..127M} presented upgraded models that take into account the changes in the electron spectral distribution at the hotspot as due to the leakage time of non-thermal particles from the acceleration site to the lobes. In the former work, the authors showed that these changes in the spectral index would induce a significantly fast drop in the lobe luminosity, when compared to the case in which this effect is neglected. In the second work, the authors showed that a detachment of the hotspot dimensions from the lobe evolution \citep[in contrast with the imposed self-similarity by][]{1997MNRAS.286..215K} would imply even faster drops in radio luminosity as forced by the strong adiabatic losses from the acceleration site to the lobes. However, a flatter evolution of luminosity, more similar to that given by KDA, is recovered if reacceleration processes take place in the lobes. Actually, \citet{2003PASA...20...94P} and \citet{2008ApJ...687..141K} showed that the hotspot size in FRII sources departs from self-similarity beyond the inner kiloparsecs, but still shows correlation with the size of the source. 

In our case, we have not used the self-similar approximation, but, by following the KDA approach, we assume that particles drift from the hotspot to the lobe at a constant rate. Then, keeping a constant spectral index, our model implicitly assumes that reacceleration at the lobes is enough to avoid a significant steepening, at least at the low-frequency range. As shown by \citet{2002A&A...391..127M}, the inclusion of reacceleration results in $P-D$ tracks similar to those obtained in KDA. Nevertheless, we plan to include a more detailed model for the non-thermal particle distribution in future work.  

\citet{2018MNRAS.475.2768H} presented an analytical model based on numerical simulations. As in our case, this model is not self-similar. A relevant difference between both works arises from the characterization of the ambient medium: the ambient pressure values at the galactic core differ by one or two orders of magnitude, being larger in our case. Finally, when estimating the source luminosity, we use the KDA model, whereas in \citet{2018MNRAS.475.2768H} the author uses an approach that does not require the time integral of particle injection and cooling. Adiabatic cooling is included via a correction to the initial estimate of the luminosity.

These differences have an impact on the resulting evolution as computed by the two models. Interestingly, although the ambient pressure is significantly smaller in the hosts considered by \citet{2018MNRAS.475.2768H} than in our case in the inner galaxy kiloparsecs, overall radio galaxies seem to propagate more slowly. This is probably due to the larger core radii in that work, which force the radio galaxy to evolve through a constant or slowly decreasing density medium for longer distances. Regarding the radio luminosity, we obtain an initial increase through the inner kiloparsec followed by a decrease until the group/cluster core is reached, at tens of kiloparsecs, where luminosity increases. After this point, depending on jet power and host properties, the luminosity can either fall or mildly increase. In contrast, all the radio galaxies in \citet{2018MNRAS.475.2768H} show a growing trend up to very large scales. This is probably caused by the properties of the hosts (namely, the values for $\beta$ considered, also in our case, as we have shown in Fig.~\ref{lumDbetabb2}), as the trend is present independent of the jet power.

\citet{2015ApJ...806...59T,2020MNRAS.493.5181T} presented a detailed evolution model. This work includes the cocoon evolution of FRI sources by taking into account a transonic phase and subsequent expansion at the sound speed, and introducing a multi-power-law pressure profile. Furthermore, the cocoon geometry is obtained via independent evolution of the working surface and lateral shock, and also estimating mixing with the shocked ambient gas through the growth of Rayleigh-Taylor instabilities. In the first of the two papers, the authors use the KDA approach to compute the radio luminosity. They report that the inclusion of the subsonic phase significantly changes the evolution of radio luminosity beyond 1~Myr. In their Fig.~3, one can observe that the purely supersonic evolution gives an increasing trend in luminosity up to large sizes, similar to the case of \citet{2018MNRAS.475.2768H}. The inclusion of a subsonic phase results in an increase in luminosity during the transonic phase, prior to the fast drop well into the subsonic phase. The increase is caused by the slower decline of the cocoon pressure once the equilibrium with the ambient pressure has been achieved (the former is forced to fall with the latter), as compared with the preceding phase of supersonic cocoon expansion. Our model also shows this behaviour for sources in which the cocoon reaches pressure equilibrium while the hotspot may still be supersonic (see the $10^{35}$ and $10^{37}$~W cases in Figs.~\ref{pressure} and \ref{lumt}).

The main difference between \citet{2015ApJ...806...59T,2020MNRAS.493.5181T} and our model is the precise calculation of cocoon geometry and the consideration of mixing that the authors of those works perform. However, because adiabatic cooling is the dominant process in determining the volume of the emitting region, we assume that mixing plays a limited role in its evolution.

In \citet{2023MNRAS.518..945T} the authors introduce the role of the jet and the calculation of the radio luminosity by means of the injection of Lagrangian particles in their modeling. The model is aimed to reproduce the results derived from numerical simulations and shows remarkable success. In contrast, our aim is to produce a simplified version that allows us to run statistical studies of radio galaxy evolution and power distributions (Beltr\'an-Palau et al., in preparation). The ambient medium profile used by \citet{2023MNRAS.518..945T} results in length evolution tracks that show deceleration beyond 1~Myr, which we only observe for the densest cores considered (see Fig.~\ref{lengthrho0a}). The core densities used in \citet{2023MNRAS.518..945T} are typically smaller than ours by several orders of magnitude, which also impacts on the values obtained for the hotspot pressure (around $10^{-6}$~Pa in our case, see Figs.~\ref{pressure} compared to $\sim 10^{-8}$~Pa, see their Fig.~8). Regarding the luminosity evolution, the jet contribution seems to play a relevant role in their case (see Figs.~\ref{lumt} and \ref{LumvsDobs} and their Fig.~9). We also introduced a simplified calculation of the jet luminosity in our model, but it resulted in negligible luminosity. This difference could be explained by the higher frequency they use (1.4~GHz in their case vs 150~MHz in our case).

\subsection{Caveats}

Our model for the evolution of radio sources represents an improvement in several aspects with respect to previous models: we include the effect of loading of thermal matter in the cocoon and extend the evolution up to the final, transonic phase, allowing us to consider the whole range of radio galaxy powers. However, our aim to make the calculations efficient enough to be applied to a large number of sources implies several simplifications: the shocked volume is approximated to a cylinder, when it should be an ellipsoid; the ambient medium is assumed to be smooth and stationary; the cocoon pressure is taken as a single value corresponding to the instant's post-shock value for the whole volume although there may be inhomogeneities, and we only implicitly take the effects of reacceleration into account. Despite these caveats, we show that the prediction power of the model is high when we compare the computed trajectories to those obtained from our simulations. This encourages us to use it for population studies of radio galaxies, a work that will be presented in a follow-up paper (Beltr\'an-Palau et al., in preparation).

Regarding the estimate of radio luminosity, we have used the simplified KDA approach, which assumes a continuous and homogeneous leaking of non-thermal particles to the lobes.
However, we also observe that this approach mimics, at least qualitatively, the role of acceleration of particles in radio lobes, which is expected to take place via turbulence and shear. Furthermore, the luminosity of the transonic phase is computed in a similar way, by simply assuming that the volume elements of emitting particles expand from the dissipation, acceleration region \citep{2014MNRAS.437.3405L} along the jet, to the transonic plumes that constitute the radio lobes in decelerated sources. This approach substitutes the consideration of expansion from the acceleration site at the hotspot to the lobes for active, supersonic sources.

\subsection{Implications on the age of radio galaxies}

Although we have seen in the previous sections that the ambient medium plays a relevant role in the evolution of radio galaxies and that there may be different galactic and cluster density/pressure profiles, we can use our results to study the implications at least for those galactic/cluster profiles that resemble the one used in this work. 

In this respect, Fig.~\ref{LumvsDobs} shows isochronal lines at different ages. We see that powerful sources approach the 500~Myr isochrone, whereas low power radio galaxies cross the detectability threshold at decreasing ages, ranging from 100~Myr to less than 50~Myr.Taking into account projection effects and the possible changes in the properties of the host galaxies, we can only expect this value to be representative of jet activity duration. We also recall that the surface brightness limitations of the LoTSS sample play a role in defining the limits of the source distribution in Fig.~\ref{LumvsDobs} \citep[see][]{2019A&A...622A..12H}. In our forthcoming paper (Beltr\'an-Palau et al., in preparation), we discuss these aspects in more detail.

Regarding large scale sources, we observe that, if undisturbed and propagating through a dilute or decreasing pressure environment, they can rapidly propagate to very large distances (reaching 1~Mpc in 100~Myr), as recently reported in \citet{2024Natur.633..537O}. However, this should still be tested via dedicated numerical simulations.

\section{Summary}
\label{conclusions}

In this paper we present an improved version of the eBC dynamical model for radio galaxy evolution \citep[see Sec.~\ref{Dynamics}][]{2002MNRAS.331..615S,2007MNRAS.382..526P} which extends the model beyond the strong-shock expansion phase, hence allowing us to consider both the very long-term source evolution and the whole range of radio galaxy powers. Besides that, the model includes the effect of loading of thermal matter in the cocoon, which is expected to occur through the entrainment of stellar and gas material in the jet within the central region of the host galaxy. The radio luminosity of the radio galaxies along their evolution, adapted from the approach in \citet{1997MNRAS.292..723K}, is described in sec.~\ref{Rlum}. Our model is aimed to be used to compute the evolution and radiative properties of large numbers of radio galaxies to perform statistical comparisons with observational samples including thousands of these objects.

In sec.~\ref{Discussion}, we have compared our model with those existing previously. A clear path of improvement would be by implementing a more detailed calculation of the evolution of non-thermal particles, or taking the jet dynamics (mass-load) into account in the determination of the jet head velocity. However, regardless of these potential improvements, this model, inspired by our experience on numerical simulations of relativistic, extragalactic jets, reproduces the overall evolution of the radio sources derived from those long, computationally demanding simulations remarkably well (see sec.~\ref{results}). Despite the expected variety of environments and situations that jets may find along their evolution, the results presented here allow us to rely on this model to study radio galaxy populations. The aim is to put constraints on, e.g., the radio galaxy power distribution or other relevant parameters that can shed light on the role that galactic activity has played along the evolution of the universe. This study is left for our next work.

\begin{acknowledgements}
This work has been supported by the Astrophysics and High Energy Physics program supported by the Spanish Ministry of Science  and Generalitat Valenciana with funding from European Union NextGenerationEU (\texttt{PRTR-C17.I1}) through grant \texttt{ASFAE/2022/005}, by the Spanish Ministry of Science through Grants \texttt{PID2022-136828NB-C43}, \texttt{PID2022-138855NB-C33} and by the Generalitat Valenciana through grant \texttt{CIPROM/2022/49}.
\end{acknowledgements}

\bibliographystyle{aa.bst}
\bibliography{biblio}

\appendix 
\section{Asymptotic limit of the spectral index}
\label{assymptotic}

As shown in Figure \ref{spectral}, the evolution of the spectral index converges at late times (when the energy losses by inverse Compton dominate over the synchrotron ones) to the value $\alpha=p/2$. This appendix is devoted to give a mathematical rationale of this result.

To make the mathematical developments simple, let us approximate the integral leading to the cocoon's luminosity $L_{\nu,c}$, \eqref{lum_lobe}, by the trapezoidal rule:
\be
\int_a^b f(x) \, dx \approx \frac{1}{2} (b-a) (f(a)+f(b))\,.
\ee
The use of such an approximation is justified because, at large times, the approximated luminosity obtained with this rule gives an almost constant fraction of the exact value for a wide range of jet powers. For instance, at $t = 1000$ Myr, if $\bar{L}_{\nu,c}$ is the luminosity of the cocoon at frequency $\nu$ as approximated by the trapezoidal rule, we obtain $\bar L_{\nu,c}\approx 0.6 L_{\nu,c}$ for all jet powers between $10^{35}W$ and $10^{40}W$. This is caused by the shape of the function inside the integral in \eqref{lum_lobe} not depending strongly on the jet power. 

Let us then look for the dependence on $\nu$ of the source luminosity at long times ($\sim 1000$ Myr) using the approximated value, $\bar L_{\nu,c}$. According to the definitions of $t_{min}$ and $\tilde \gamma$ (see the text following eq.~\ref{gamma}), $\tilde\gamma(t_{min}(t),t)\to\infty$ for any given time $t$. Therefore, the function inside the integral in \eqref{lum_lobe} vanishes at $t_i=t_{min}(t)$. Furthermore, by the definition of $\tilde \gamma$, $\tilde\gamma(t,t)=\gamma(t)$. Then, we obtain
\be\label{lum_approx}
\bar L_{\nu, \, c}(t)=(t-t_{min}(t))\frac{4 L_0\sigma_{\mathrm{T}}c}{6 \nu(k(t)+2)} n_0(t) \gamma(t)^{3-p} \left(\frac{p_h(t)}{p_c(t)}\right)^{-1+\frac{1}{\Gamma}}\, .
\ee

In addition, $t_{min}(t)$ can be estimated analytically under some assumptions. At late times ($\sim 1000$ Myr) the energy density of the CMB radiation dominates over the magnetic energy density, i.e., $u_C \gg u_B$. Taking this into account and assuming that the cocoon's pressure decreases with time as $t^b$ ($b<0$), the differential equation \eqref{edo_gamma} can be integrated analytically. We then calculate the value of $t$ at which $\tilde \gamma$ tends to infinity, and obtain 
\be
t_{min}(t)=t^{\frac{b}{b+3\Gamma}}\left(t-\frac{m_e c(b+3\Gamma)}{4\sigma_T u_C\Gamma\gamma(t)}\right)^{\frac{3\Gamma}{b+3\Gamma}}\,.
\ee
A further simplification can be applied considering that, according to \eqref{gamma}, the required Lorentz factor $\gamma(t)$ for the emission at frequency $\nu$ increases with time as $t^{-b/4}$, so we can approximate the previous expression at late times by expanding it at first order in $1/\gamma(t)$
\be
t_{min}(t)=t-\frac{3m_e c}{4\sigma_T u_C\gamma(t)}+O\left(\gamma(t)^{-2}\right)\, .
\ee
Substituting in \eqref{lum_approx} we obtain
\be\label{lum_approx2}
\bar L_{\nu, \, c}(t)=
\frac{L_0\, m_e\, c^2}{2\,u_C\,\nu\,(k(t)+2)} n_0(t) \,\gamma(t)^{2-p} \left(\frac{p_h(t)}{p_c(t)}\right)^{-1+\frac{1}{\Gamma}}+O\left(\gamma(t)^{1-p}\right)\, .
\ee

Finally, since according to \eqref{gamma}, the Lorentz factor depends on the frequency as $\gamma(t)\propto\nu^{1/2}$, it is easy to see that the cocoon's luminosity behaves as $L_{\nu,c}\propto\nu^{-p/2}$. And therefore, from \eqref{spectral} we conclude that $\alpha$ tends to $p/2$ when inverse Compton dominates the energy losses in the cocoon. This proof can be also derived in the same way when the lobes are in pressure equilibrium with the ambient medium, by simply changing the pressure functions.

This result, derived for the luminosity of the source at a given frequency range, implies that the electron distribution power-law changes from $p$ to $p+1$, as expected from inverse Compton losses \citep{1970ranp.book.....P}.

\end{document}